\documentclass[11pt,a4paper]{article}
\usepackage[utf8]{inputenc}
\usepackage[DIV=10]{typearea}
\usepackage{graphicx}
\usepackage{url}
\usepackage{multirow}
\usepackage{longtable}
\usepackage{caption}
\usepackage{enumitem}
\usepackage{titlesec}
\usepackage{hyperref}
\titleclass{\subsubsubsection}{straight}[\subsection]
\newcounter{subsubsubsection}[subsubsection]
\renewcommand\thesubsubsubsection{\thesubsubsection.\arabic{subsubsubsection}}
\titleformat{\subsubsubsection}
  {\normalfont\normalsize\bfseries}{\thesubsubsubsection}{1em}{}
\titlespacing*{\subsubsubsection}
{0pt}{3.25ex plus 1ex minus .2ex}{1.5ex plus .2ex}
\makeatletter
\renewcommand\paragraph{\@startsection{paragraph}{5}{\z@}%
  {3.25ex \@plus1ex \@minus.2ex}%
  {-1em}%
  {\normalfont\normalsize\bfseries}}
\renewcommand\subparagraph{\@startsection{subparagraph}{6}{\parindent}%
  {3.25ex \@plus1ex \@minus .2ex}%
  {-1em}%
  {\normalfont\normalsize\bfseries}}
\def\toclevel@subsubsubsection{4}
\def\toclevel@paragraph{5}
\def\toclevel@paragraph{6}
\def\l@subsubsubsection{\@dottedtocline{4}{7em}{4em}}
\def\l@paragraph{\@dottedtocline{5}{10em}{5em}}
\def\l@subparagraph{\@dottedtocline{6}{14em}{6em}}
\makeatother
\title{Proposal for the ILC Preparatory Laboratory (Pre-lab)}
\author{International Linear Collider \\ International Development Team }
\date{1 June 2021}

\begin{document}
\maketitle
\vspace{3cm}
\begin{abstract}
\noindent 
During the preparatory phase of the International Linear Collider (ILC) project, all technical development and engineering design needed for the start of ILC construction must be completed, in parallel with intergovernmental discussion of governance and sharing of responsibilities and cost. The ILC Preparatory Laboratory (Pre-lab) is conceived to execute the technical and engineering work and to assist the intergovernmental discussion by providing relevant information upon request. It will be based on a worldwide partnership among laboratories with a headquarters hosted in Japan.  This proposal, prepared by the ILC International Development Team and endorsed by the International Committee for Future Accelerators, describes an organisational framework and work plan for the Pre-lab.  Elaboration, modification and adjustment should be introduced for its implementation, in order to incorporate requirements arising from the physics community, laboratories, and governmental authorities interested in the ILC.
\end{abstract}
\newpage
\section*{Preface}
This is a proposal for the Preparatory Laboratory (Pre-lab) for the International Linear Collider (ILC). It has been prepared by the Executive Board (EB) of the ILC International Development Team (IDT) with contributions from the three IDT working groups: Working Group 1 (WG1) for function and organisational structure of the Pre-lab, Working Group 2 (WG2) for accelerator and facilities, and Working Group 3 (WG3) for physics and detectors. The IDT was established by the International Committee for Future Accelerators (ICFA) with a mandate to prepare the Pre-lab as a preparatory phase of the ILC project. The EB members were appointed by ICFA and the working group members by the EB. 

The document has been endorsed by the ICFA and outlines the organisational framework, an implementation model and work plan of the Pre-lab. It provides information to the laboratories and governmental authorities interested in the ILC project to allow them to consider participation. 

Further details will be developed during the actual implementation process of this proposal. The implementation will reflect input from governmental authorities in Japan and elsewhere, from laboratories that are the basis of the Pre-lab’s collaborative work, and from the international physics community that is the driving force for the ILC project.

\vfill
\noindent
\makebox[\textwidth][r]{
\begin{minipage}{.93\textwidth}%
The members of the ILC International Development Team Executive Board are:
\\
\begin{tabular}{ll}
\hspace*{3mm}T.~Nakada (EPFL); & Chair of WG1 and EB Chair
\\
\hspace*{3mm}S.~Michizono (KEK); &Chair of WG2
\\
\hspace*{3mm}H.~Murayama (UCB \& University of Tokyo); &Chair of WG3
\\
\hspace*{3mm}A.~Lankford (UCI); &Representing Americas
\\
\hspace*{3mm}G.~Taylor (University of Melbourne); &Representing Asia-Pacific 
\\
\hspace*{3mm}S.~Stapnes (CERN): &Representing Europe 
\\
\hspace*{3mm}Y.~Okada (KEK); &KEK liaison 
\\
\end{tabular}\\
\noindent assisted by T.~Tanabe (Iwate Prefectural University) as Scientific Secretary and R.~Takahashi (KEK) for communication.  Working group members can be found on the ILC IDT web site\footnote{WG1 \url{https://linearcollider.org/team/wg1/}, WG2 \url{https://linearcollider.org/team/wg2/}, WG3 \url{https://linearcollider.org/team/wg3/}.}.
\end{minipage}
}
\newpage
\tableofcontents
\newpage
\section{Introduction}
The International Linear Collider (ILC) is a large-scale scientific facility under development for research in particle physics. Its purpose is to produce collisions of high-energy beams of electrons and positrons with center-of-mass energy of 250~GeV~\cite{Evans:2017rvt}. It will provide a well-characterized initial state to study interaction of elementary particles at energies typical of the environment only a trillionth of a second after the Big Bang. The initial focus will be to understand the properties of the newly discovered Higgs boson to great precision, which is believed to point to physics beyond the Standard Model of particle physics. At the same time, it will search for dark matter, study the stability of the Universe, look for clues of unification of forces and matter, and address many other fundamental scientific questions. The linear design of the ILC allows for extension in the future to reach higher collision energies. The ILC can also host additional experiments with extracted beams, at the beam dump, and near the collision point.

\subsection{Brief history of the ILC}\label{subsection:history}
There is a long history of development of the ILC. The need for a linear collider was recognized already in the 1960's~\cite{LC:1965} given the energy loss due to unavoidable synchrotron radiation by beams in circular colliders. To achieve power-efficient acceleration, the development of superconducting radio frequency (SRF) cavities started in earnest in the 1980's. Over four decades, intensive research and development achieved much higher acceleration gradients and reduced costs of SRF by more than an order of magnitude. SRF provides better tolerance compared to room-temperature klystron-based radiofrequency designs, and SRF was selected as the technology of choice in 2004 by the International Technology Recommendation Panel \cite{ITRP:srf} chaired by Barry Barish (2017 Nobel Laureate in Physics). The International Committee for Future Accelerators (ICFA), a body created by the International Union of Pure and Applied Physics in 1976 to facilitate international collaboration in the construction and use of accelerators for high energy physics, recommended launching the Global Design Effort (GDE) to produce a Technical Design Report (TDR) for the ILC as an international project. The GDE, led by Barish, successfully produced the TDR in 2013~\cite{Behnke:2013xla, Baer:2013cma, Adolphsen:2013jya, Adolphsen:2013kya, Behnke:2013lya} in a purposely site-independent fashion.

The scientific merit of the ILC has long been recognized. The energy scale of the weak interaction, which makes the Sun burn and which synthesized the chemical elements in cosmic history, was pointed out to be around 250~GeV as early as 1933 by Enrico Fermi. The need to reach this energy scale has been accepted since then, but the precise target energy was not clear. Early discussions for linear colliders called for 1500~GeV as a safe choice for guaranteed science output, while the GDE focused on 500~GeV for the study of the Higgs boson based on the data in early 2000's. It was only in 2012 when the Higgs boson was discovered at the Large Hadron Collider (LHC) at CERN~\cite{Aad:2012tfa, Chatrchyan:2012ufa} that an initial  target energy for the ILC of 250~GeV became clear. In the same year, the Japan Association of High Energy Physicists (JAHEP) issued a report proposing to host the ILC in Japan with 250~GeV center-of-mass energy as its first phase \cite{JAHEP:2012}. The European Strategy for Particle Physics updated in 2013 \cite{CERN-ESPP:2013} highlighted ``{\it the ILC, based on superconducting technology, will provide a unique scientific opportunity at the precision frontier.}'' The 2014 report of the US Particle Physics Project Prioritization Panel (P5)  \cite{ParticlePhysicsProjectPrioritizationPanel(P5):2014pwa}, citing the 2012 discovery of the Higgs boson, identified ``{\it Use the Higgs boson as a new tool for discovery}'' as a science driver for particle physics and stated ``{\it As the physics case is extremely strong, all (funding) Scenarios include ILC support}''.

Intense discussions ensued worldwide about how to realize the ILC. The Japanese Ministry for Education, Culture, Sports, Science, and Technology (MEXT) asked  
the Science Council of Japan (SCJ) to look into the scientific case and socioeconomic merit of hosting the ILC in Japan as well as its technological feasibility, costs, and management structure~\cite{MEXT-SCJ:2013}. MEXT then formed its own ILC Advisory Panel\footnote{MEXT also contracted an external assessment on the project management, risk, and technical and economical merit to the Nomura Research Institute (\url{https://www.nri.com/en}}
that ran from 2014 to 2018 with four subgroups~\cite{MEXT1}, which also reassessed the new baseline of 250~GeV~\cite{MEXT2}. After this, MEXT asked again SCJ to reevaluate the case for the 250~GeV baseline~\cite{MEXT-SCJ:2018}. 

The US government expressed an interest in engaging in discussions with the Japanese government on governance and preparatory activities towards the ILC being hosted in Japan~\cite{ICFA:2021}.
The 2020 update of the European Strategy for Particle Physics \cite{CERN-ESPP:2020} stated ``{\it An electron-positron Higgs factory is the highest-priority next collider}'' and added ``{\it The timely realisation of the electron-positron International Linear Collider (ILC) in Japan would be compatible with this strategy and, in that case, the European particle physics community would wish to collaborate.}'' From the scientific standpoint, it is highly valuable to have the ILC start taking data while High-Luminosity LHC is still in operation. 

ICFA chartered the International Development Team (IDT) in August 2020 \cite{ICFA2020} with a charge to prepare for the creation of the Pre-lab as a preparatory phase for the ILC construction. IDT is hosted by KEK, the national laboratory for high-energy accelerators in Japan.
\subsection{Mandate of the Pre-lab}\label{subsection:mandate}
The main purpose of the Pre-lab is to bring the technical and engineering work of the ILC project to a point where the construction can be started. Although the high energy physics community has considerable experience in constructing large accelerators and much technical work has already been completed for the TDR by the GDE, further effort is still required to be ready for construction. For the civil engineering work, a site specific study must be conducted, which was not possible for the TDR. These efforts will lead to a more accurate cost estimate of the ILC project. An equally important task is to ensure an inspiring ILC physics programme.

During the Pre-lab phase, it is expected that government authorities of interested nations are forging agreements on the sharing of the cost and responsibilities for the construction and operation of the ILC facility 
and on the organisational structure and governance of the ILC Laboratory. The government authorities may wish the Pre-lab to provide information on resource and technical matters during their intergovernmental discussion. 

With this consideration, the ILC Pre-lab will address the following topics:\\[2mm] 
\noindent
{\bf Completion of technical preparations and production of engineering design documents for the accelerator complex. } \\ \noindent 
While the GDE resolved most of  technical details, as elaborated in the TDR, there are some items which require further study, such as the positron source and the beam dumps. Some of these open issues were also pointed out by the expert panel of the Japanese Ministry of Education, Culture, Sports, Science and Technology (MEXT) for the ILC and by the Science Council of Japan Deliberation Committee. Since the completion of the TDR almost ten years ago, R$\&$D has produced significant improvements in some items, such as the superconducting radio frequency cavities, and the ILC design must take such improvements into account. Furthermore, in order to be ready for construction, engineering design specifications must be made and documented. 

 The work of the Pre-lab will be organised into work packages. Categorisation of work packages will allow laboratories wishing to participate in the Pre-lab to identify work to match their interest, expertise, and resources. Reference costs and human resource requirements are also needed for use by the laboratories in preparing their resources in order to make in-kind contributions.  
\\[2mm]%
{\bf Compilation of design studies and documentation of the civil engineering and site infrastructure work, and of the environmental impact assessment. }\\ 
\noindent 
In the TDR, no specific site was assumed for the project implementation. Proper technical preparation for construction can only be made for a specific site with necessary adjustments required by the geological constraints of the site. It should cover the civil engineering for the accelerator complex, office and other buildings for the ILC Laboratory, as well as infrastructure such as electricity, communication network, water supply and waste management within the site. Design documents must then be prepared for the construction. 

An environmental impact assessment for the ILC construction, operation and dismantling for the candidate site is required. In parallel, intensive communication with the local community must be initiated before the governmental decision on the site.  
\\[2mm]%
{\bf Community guidance to develop the ILC physics programme that will fully exploit its potential. }\\ 
\noindent 
The Pre-lab must pave the way for the ILC laboratory to set up its physics programme by encouraging and guiding the community to propose a wide range of experiments that could exploit the full physics potential of the ILC. The Pre-lab needs to provide clear guidelines and a time frame for the community to develop ideas for ILC experiments and to support the development efforts. 
\\[2mm] %
{\bf Provision of information to national authorities and to Japanese regional authorities to facilitate development of the ILC Laboratory.} 
\\%
\noindent 
The Pre-lab management should be ready to provide information to national authorities upon request to aid intergovernmental negotiations to set up the ILC Laboratory. Such information could include possible organisational structures and operational models or technical issues relevant for the cost-sharing discussion. Interacting with the local community and regional government of the candidate site will also be the work of the Pre-lab directorate.
\\[2mm]
{\bf Coordination of outreach and communication work.} \\ \noindent
Given the scale of the ILC project, communicating the unique scientific and societal benefits of the ILC to the broader community of scientists, general public, policy makers, and government authorities worldwide is of vital importance. Outreach and communication activities must be accomplished as a common effort by the laboratories participating in the Pre-lab coordinated by the Pre-lab management. The management will also be responsible for establishing a coherent strategy for this effort. 

\subsection{Principle of Pre-lab operation}
The Pre-lab will be organised as an international collaboration of laboratories worldwide. The laboratories could be national laboratories, intergovernmental laboratories, such as CERN, or university laboratories. All the technical preparation and engineering work will be organised as work packages and will be provided as in-kind contributions by the laboratories. Laboratories' contributions will be formalised through the exchange of Memoranda of Understanding (MoUs). 

An assembly of representatives from the participating laboratories (the Steering Board in Section~\ref{section:organisation}), is the highest decision-making body of the Pre-lab. An 
assembly of funding authorities (the Committee of Funding Authorities in Section~\ref{section:organisation})
is a forum for funding agencies and national authorities who support the participating laboratories to monitor the progress of the Pre-lab activities and provide advice when needed. The Pre-lab management consists of a director and associate directors, referred to as the Directorate in Section~\ref{section:organisation}, supported by a small team, forming the ``Central Bureau'' (described in Section~\ref{section:organisation}), to be located in Japan. While the director will have the overall responsibility and lead the management, associate directors will have well-defined responsibilities in accelerator, civil engineering and infrastructure, and research. The management will represent and operate the Pre-lab with overall coordination of the Pre-lab work. 

Execution of in-kind contributions including their funding is the responsibility of the laboratories who sign the various MoUs and will be fully managed by them. However, any change to the scope defined in the MoUs must be discussed and agreed to by the Pre-lab as a whole in the Steering Board. MoUs will be drafted by the IDT Executive Board in discussion with the laboratories. After the start-up of the Pre-lab, drafting of the MoUs will be taken over by the Pre-lab management. 

Operation of the Central Bureau, including the employment of its personnel, will take place in Japan. Some  specialists in the support groups may be temporarily relocated from participating laboratories as in-kind contributions to the Pre-lab. In order to ensure neutrality, the director and other members of the management should be paid through cash contribution from the participating laboratories to a central fund, while the remaining cost for the operation will be covered by Japan. The responsibility for the civil engineering and site-related work will be taken by Japan, where the machine will be located. The responsibilities for the accelerator work will be shared, as in-kind contributions, more or less equally among the three regions: the Americas (mainly the US), Asia (mainly Japan), and Europe. 
\section{Pre-lab organisation}\label{section:organisation}
The proposed organisation and governance structure for the ILC Pre-lab has evolved from the report of the KEK International  Working Group on the ILC Project in September 2019 \cite{KEKInternationalWorkingGroup:2019spu}. The proposal also takes note of the experience of the large LHC experiment collaborations, as well as the organisational structure of CERN. It should be noted that the future ILC organisation may well evolve from that of the Pre-lab. However, considering the difference in aims and complexity, such an evolution is not a key driver of the proposed Pre-lab organisation. The proposed Pre-lab organisation takes account of both the highly distributed nature of the work to be carried out and the finite duration of the Pre-lab operation. 
\subsection{Organisation Structure}
\begin{figure}[tbh]
\centering
\includegraphics[width=\textwidth]{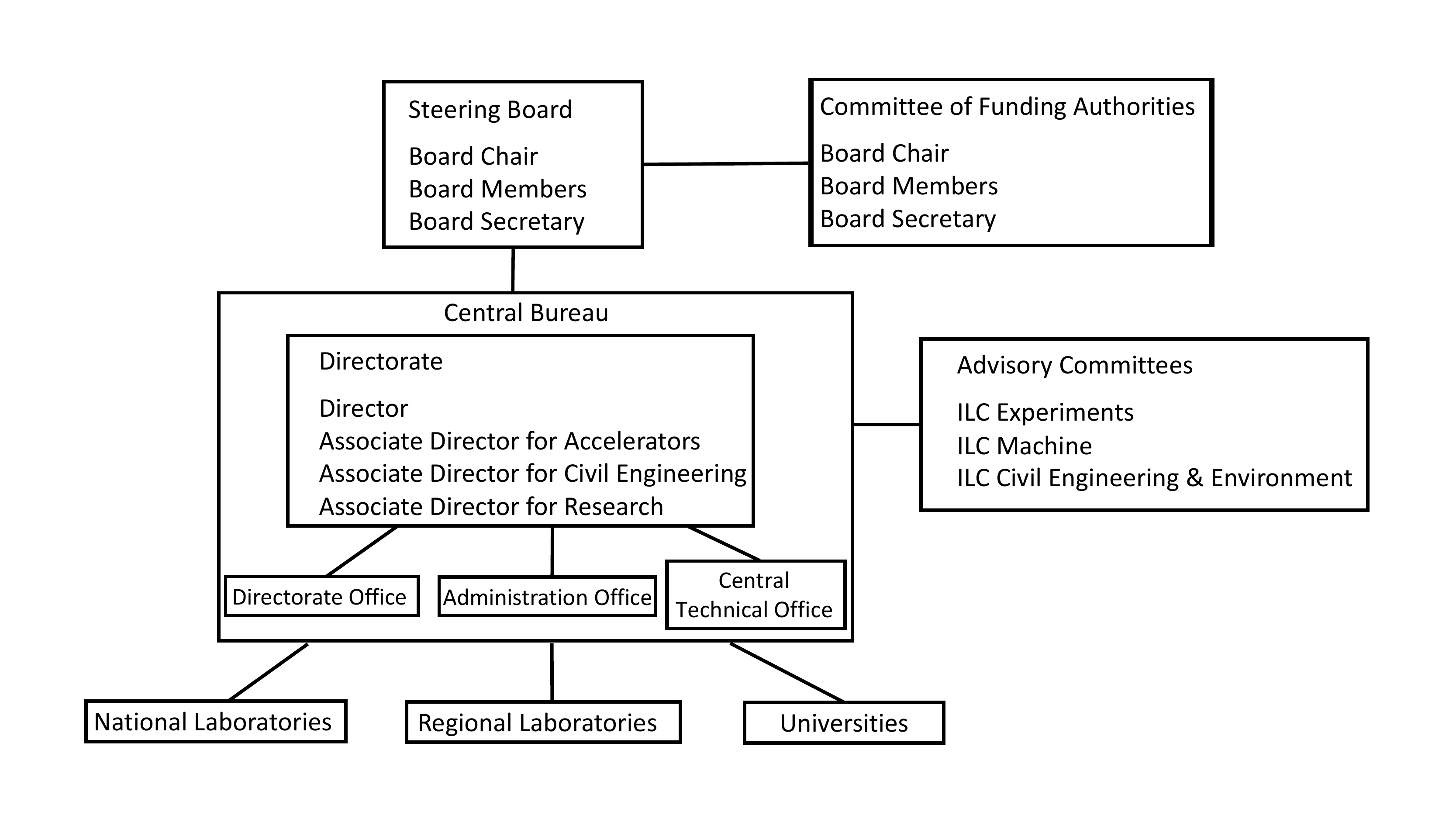}\vspace{-9mm}
\caption{Pre-lab Organisation Chart}\label{fig:Org-Chart}
\end{figure}

The proposed organisation is shown schematically in Figure~\ref{fig:Org-Chart}. 
The roles of the various groupings in the figure are described below.
\subsection{Steering Board}
The Steering Board is the highest decision making body of the Pre-lab. It appoints the Pre-lab Director. The Steering Board approves Associate Directors recommended by the Director, who will consult with appropriate regional bodies before making such recommendations.
It also approves new members of the Pre-lab. 
The Steering Board receives reports from the Pre-lab Director on the progress and status of the project. 
It makes major decisions on Pre-lab activities and budget, as well as any changes to the distribution of responsibilities for the in-kind contributions.
\newline
\textbf {Membership:} Chair (elected by the Steering Board from its members), laboratory directors or equivalent (or their delegates) of those Pre-lab Members (see Subsection~\ref{subsection:members}) providing in-kind contribution and other resources to the Pre-lab above a minimum level. New membership applications can be reviewed at any meeting.
\newline
\textbf {Suggested Meeting frequency:} approx. four times per year
\newline
\textbf {Attendance:} Pre-lab Management; Board Secretary; and other members of Central Bureau and chairs of the Advisory Committees as requested for particular items.
\subsection{Committee of Funding Authorities}
The Committee of Funding Authorities provides a forum for funding agencies and national authorities of Pre-lab Members providing in-kind contribution and other resources to the Pre-lab above a minimum level to monitor the progress of the Pre-lab activities. 
The Committee of Funding Authorities can provide feedback on progress and activities to the Steering Board. Requests to the Steering Board can be made by the Committee of Funding Authorities Chair through the Steering Board Chair. It could provide an interface to FALC (Funding Agencies for Large Colliders).
\newline
\textbf {Chair:} Representative of Japanese national authority.
\newline
\textbf {Members:} Funding agency or national authority representatives of those Pre-lab Members represented in the Steering Board, Chair of the Steering Board.
\newline
\textbf {Meeting frequency:} Once or twice a year.
\newline
\textbf {In attendance:} Pre-lab management, Committee Secretary.
\subsection{Central Bureau}
The Central Bureau is a compact headquarters for the ILC Pre-lab management team located in Japan. It consists of Directorate, Directorate Office, Central Technical Office and Central Administration Office. It will be a legal entity under Japanese law. 
\subsubsection{Directorate}
The Directorate will have the collective responsibility for the overall running of the Pre-lab. 
More detail for their mandate is given in the following sections. 
\newline
\textbf{Members:} Director, Associate Directors 
\newline
\textbf{In attendance:} Head of Administration Office, who acts as Secretary
\newline
\textbf {Suggested Meeting frequency:} approx. weekly 
\vspace{-0.2cm}
\subsubsubsection {Director}
The Director has overall responsibility for delivering the Pre-lab project.
The Director will take guidance from the participating laboratories on matters that impact upon their resources and expertise.
Success in achieving an intergovernmental agreement for the ILC depends critically on the close interactions of laboratories participating in the Pre-lab with their government authorities.
The Director will play a leading role in facilitating such interactions.
As inter-governmental negotiations are expected to cover sharing of responsibilities and work packages, input on technical considerations and capacities will be required. The Director will be the point of contact for such input and will provide necessary information to assist in intergovernmental negotiations including the provision of proposed governance schemes for the ILC if so requested.
\subsubsubsection {Associate Director for Accelerators}
Key roles include: project management and coordination of in-kind contributions and work packages for the ILC Pre-lab accelerator activities and the ILC engineering design.
The Associate Director will prepare the process by which collaborators can bid for Work Breakdown Structure (WBS) items and will be responsible for maintaining a WBS list of in-kind contributions for the ILC construction.
\subsubsubsection {Associate Director for Civil Engineering}
Key roles include: project management of the work to complete the ILC civil engineering design, and interface with environmental investigations and local government.
\subsubsubsection {Associate Director for Research}\label{subsection-research-dir}
Key roles include: launching the various stages of approval for the ILC experimental programme, as described in  Subsection~\ref{sec:physicstimeline}, organisation of common tasks and resources for computing and detectors including potential work packages, to foster appropriate meetings, workshops and conferences, and to issue documents as appropriate to enhance understanding and appreciation of the physics of the ILC.
\subsubsection{Directorate Office}
The Directorate Office will support the Pre-lab operations including secretarial and other functions. Outreach and communications will be managed from the Directorate. It will also assist the Director and Associate Directors in interfacing with government authorities.
\newline
\textbf{Members:} Director’s executive assistant; Associate Directors’ executive assistants,  heads of sections responsible for public engagement, communication and publicity, and international and government affairs. 
\subsubsection{Central Technical Office}
A key function of the Central Technical Office is project management support for supervision and follow-up of all the necessary accelerator technical preparation and engineering work packages associated with the ILC Pre-lab, as described in Subsection~\ref{section:work_accelerator}. 

The Office will also coordinate the work to complete the remaining ILC technical preparations and to produce engineering design documentation. Technical and engineering work is expected to be broadly distributed internationally. While various levels of regional coordination of the work is expected, the Central Technical Office will be responsible for bringing together the distributed design information into the ILC engineering documentation. These functions are the responsibility of the Associate Director for Accelerators. The Central Technical Office will provide essential coordination of the activities. 

The Office will provide supervision and coordination of the IT and documentation support for the engineering design development.

The Office will provide support and coordination of the physics and detector preparatory work during the Pre-lab era, described in Subsection 4.3. These office functions are the responsibility of the Associate Director for Research. The Central Technical Office will provide essential support and coordination of the activities.
\subsubsection{Central Administration Office}
The Central Administration Office will provide assistance in a range of administrative tasks including budget tracking, travel organisation, claims reimbursement, visitor and meeting support, and IT support. 
The Central Administration Office staff will include heads of budget tracking and human resources, travel and claims, visitor and meeting support, and head of IT support.
\subsection{Pre-lab Members}\label{subsection:members}
The Pre-lab Members are regional, national, and university laboratories.
Membership is expected to be dynamic. 
Founding members will inaugurate the Pre-lab as described in  Section~\ref{section:legal-start-up}. 
Membership will follow the signing of MoUs. 
Individual MoUs  will be signed by the Pre-lab Director and the director (or equivalent) of each member laboratory, or funding agency, as appropriate. Additional members can be added through individual MoUs as laboratories become ready to do so.
\subsection{Advisory Committees}\label{subsection:committee}
Advisory committees will be established to advise the Directorate. The chairs of the committees will regularly report their findings to the Steering Board.  
\subsubsection{ILC Experiments Advisory Committee (ILCXAC)}
\textbf {Functions:} Advisory to the Associate Director for Research. The ILCXAC will follow the Pre-lab physics activities and make peer reviews of the various stages of the experimental programme approval process (see Subsection~\ref{sec:physicstimeline}). 
\newline
\textbf {Members:} approx. 10-15, appointed by the Director, including 2 cross-members from ILCMAC.
\newline
\textbf {Meeting frequency:} as required, at least 1 per year.
\subsubsection{ILC Machine Advisory Committee (ILCMAC)}
\textbf {Functions:} Advisory to the Associate Director for Accelerators. The ILCMAC will monitor the progress of work packages and review the engineering design documentation and its intermediate stages.
\newline
\textbf {Members:} approx. 10-15 members, appointed by the Director, including 2 cross-members from ILCXAC.
\newline
\textbf {Meeting frequency:} as required, at least 1 per year.
\subsubsection{ILC Civil Engineering and Environment Advisory Committee (ILCCEAC)}
\textbf {Function:} Advisory to the Associate Director for Civil Engineering. The ILCCEAC will provide advice on civil engineering plans and environmental assessment/geological surveys.
\newline
\textbf {Members:} approx. 5-10 members, appointed by the Director, including 2 cross-members from ILCXAC, 2 from ILCMAC.
\newline
\textbf {Meeting frequency:} as required, at least 1 per year.
\section{Required legal structure and Pre-lab start-up process}\label{section:legal-start-up}
\subsection{Legal structure}
The Pre-lab as a whole will be organised and governed as a collaboration of laboratories worldwide regulated through Memoranda of Understanding (MoUs). However, members of the Central Bureau will mostly be employed in the host location in Japan. Thus a Japanese legal framework would be required for the employment of the members and operation of the Central Bureau. If such a legal framework can also accommodate the whole Pre-lab organisation, it may strengthen the governance of the Pre-lab, although that is not mandatory. 

The most appropriate solution for the legal structure for this purpose appears to be a ``General Incorporated Association'' (GIA) under Japanese law. A GIA can be easily started by ``founders'', at least two of them who are natural or judicial person, registering it to an appropriate authority following Japanese law. Other members can join and govern the GIA as ``stakeholders'' through the ``General Assembly'' of the GIA together with founders. With this structure, the members of the Central Bureau can be paid as employees of the GIA. If the GIA is recognised as ``nonprofit'', no corporate tax will be charged for the income such as the government grant and common fund payment by the laboratories. 

The GIA could be only for operating the Central Bureau in Japan with a simple structure. If it were to incorporate the whole Pre-lab, an additional structure needs to be defined through ``Articles of Association''. 

The GIA must be established before the Pre-lab since it forms the legal structure under which the Pre-lab operates. Other than this, the two organisations can evolve in parallel.
\subsection{Pre-lab start-up}
The Pre-lab is conceived as a partnership of laboratories worldwide for preparing ILC construction. Establishment of the Pre-lab can start with an agreement among a few major laboratories (and/or funding agencies, as appropriate) to form the Pre-lab, and the Pre-Lab can then be enlarged by addition of other partner laboratories. In this way, a complex process of establishing an agreement among all potential partner laboratories as a prerequisite to creation of the Pre-lab can be avoided. The agreement of the founding laboratories could be made as a common declaration to commit resources to the Pre-lab program, with the approval of their funding authorities, as necessary. 

With the help of ICFA, the IDT will identify potential founding laboratories and facilitate discussion among them. Some indication that the Japanese government is moving towards expressing its interest in hosting the ILC in Japan as an international project will be necessary for discussion to progress to a concrete level, e.g. sharing of responsibilities and providing resources. It is also important that the international physics community interested in the ILC will continue communicating the importance of the ILC and their activities to the public and government authorities of their countries.  

For the founding laboratories to reach final agreement for forming the Pre-lab, it will be necessary that the Japanese government expresses its interest to host the ILC in Japan and invites partner states to discuss how the ILC can be realised as an international project. With an agreement in place, the founding laboratories will then appoint the Pre-lab Directorate, after consultation with other potential Pre-lab partner laboratories regarding suitable candidates. The legal structure in Japan, i.e. GIA, must be in place to accommodate the Pre-lab Central Bureau. With the completion of these steps, the Pre-lab will be officially established, and its enlargement can commence.

In parallel with discussion among the founding laboratories, the IDT will continue ongoing exploration of specific interests and expertise of laboratories planning to participate in the Pre-lab program. Possible interests of laboratories with relevant capabilities and expertise have already been identified for all the Pre-lab work packages. Along with the progress in the Pre-lab formation, discussion will advance to each laboratory’s specific in-kind contributions to the work packages and to the engineering design effort, described in Subsection~~\ref{subsection:WP} and \ref{subsection:EDR}, respectively. The agreed upon in-kind contributions will be documented in MoU’s, consisting of a common part describing the general terms for Pre-lab partnership and of customized appendices defining the specific in-kind contributions of the partner laboratory. 

The nascent Pre-lab will rapidly be enlarged by partner laboratories via signing of an MoU between each prospective partner and the Pre-lab Director. The funding authorities of the partner laboratories will be invited to become members of the Committee of Funding Authorities. In-kind contributions of the founding laboratories will also be finalised at this stage, and the Pre-lab can commence its full programme.
\section{Pre-lab work plan}\label{section:work}
\subsection{Accelerator}\label{section:work_accelerator}
The ILC accelerator consists of the following domains: 
\begin{enumerate} [label= \alph* ]
\item electron and positron sources,
\item damping rings (DRs) to reduce $e^-$ and $e^+$ beam emittance (a quantity corresponding to the spread of the beam),
\item the beam transportation from the damping rings to the main linear accelerators (RTML),
\item the main linear accelerators (main linacs or MLs) to accelerate the $e^-$ and $e^+$ beams using superconducting radio frequency (SRF) technology,
\item beam delivery and final focusing system (BDS) to focus and minimize the final beam size, in order to maximize luminosity, and to optimize the machine and detector interface (MDI) in the interaction region where experiments are installed, and
\item the beam dumps (Dump), where the beam ends after passing through the interaction region
\end{enumerate}
Common technologies, such as superconducting magnets and vacuum systems, are required by the various domains. Groups that support such technologies will provide specialized technical design and development to all domains requiring their expertise.

The principal accelerator activities of the ILC Pre-lab are technical preparations and engineering design and documentation.
These activities will be conducted in parallel with intergovernmental negotiations for the ILC Laboratory.

The deliverables of the Pre-lab accelerator activities, both technical preparations and engineering design and documentation, will be provided as in-kind contributions by member laboratories of the Pre-lab. A work breakdown structure (WBS) defining all Pre-lab accelerator activities is currently being developed. Overall management of worldwide Pre-lab accelerator activities will be provided by the Associate Director for Accelerators, assisted by the Central Technical Office. 
It is foreseen that the activity in each domain and on each common technology will be led by a manager drawn from one of the member laboratories.
Similarly, each technical preparation and engineering design work package will be led by a manager drawn from one of the member laboratories, guided by the domain and common technology managers. The detailed organization chart for Pre-lab accelerator activities will be defined by the Pre-lab Directorate. The ILC Machine Advisory Committee (ILCMAC), in its advisory role to the Associate Director for Accelerators, will monitor technical progress and review the engineering design and documentation.

\subsubsection{Technical preparation activities}\label{subsection:WP}
The Pre-lab technical preparation activities consist of the R\&D necessary to
\begin{enumerate} [label= \alph* ]
\item eliminate remaining technical uncertainties, including those pointed out in reviews by MEXT’s ILC Advisory Panel and the Science Council of Japan~\cite{MEXT1, MEXT2, MEXT-SCJ:2018},
\item incorporate significant technical advances in the years following the Technical Design Report, and 
\item enable completion of the Engineering Design Report and a reliable estimate of the cost and human resource requirements of the ILC Project.
\end{enumerate}

\begin{figure}[b!]
\centering
\includegraphics[width=\textwidth]{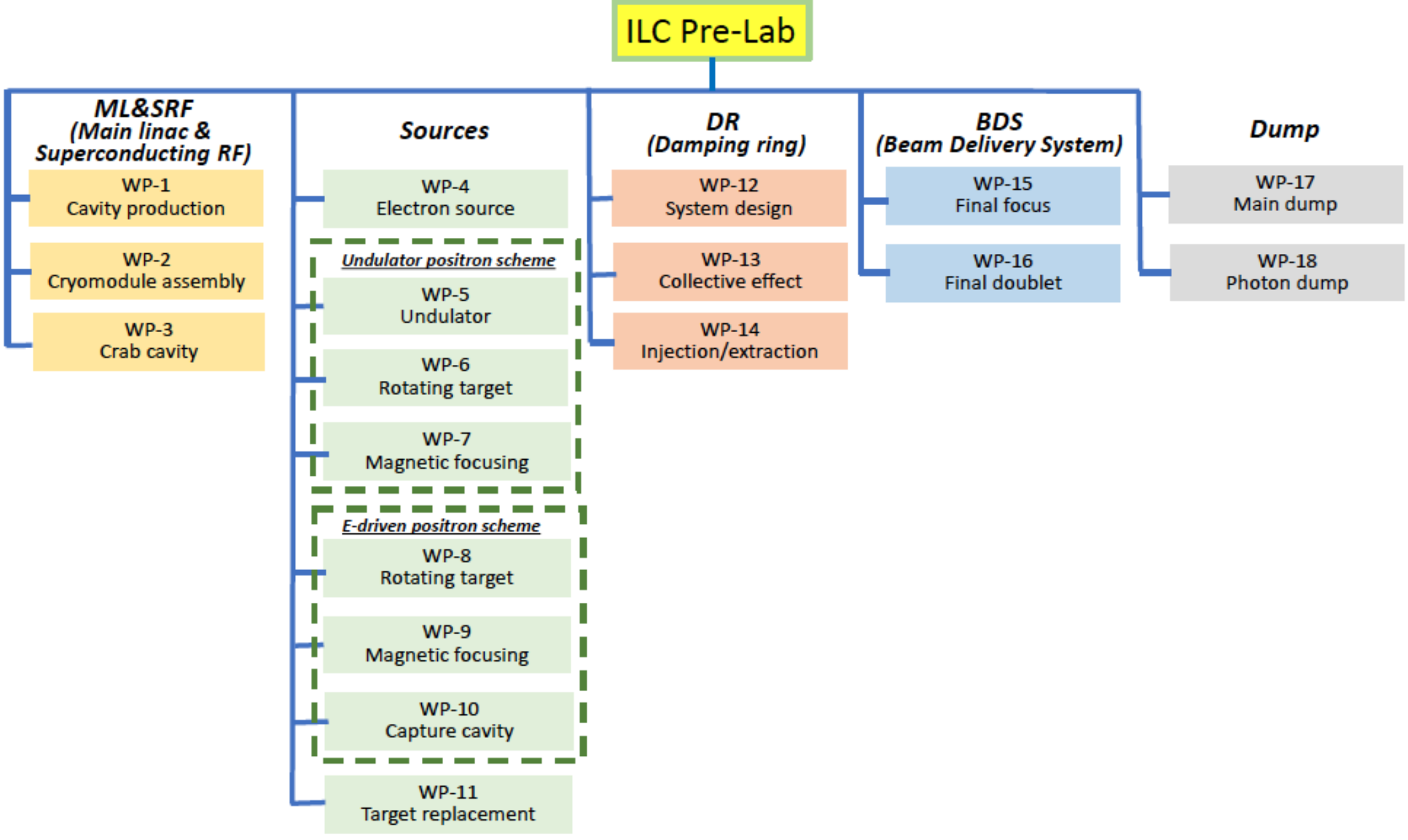}
\caption{Summary of work packages.}\label{fig:Acc-WorkPackages}
\end{figure}

A total of eighteen work packages~(WPs) over five of the accelerator domains are proposed, as illustrated in Figure~\ref{fig:Acc-WorkPackages}. There are three WPs in the Main Linac and Superconducting RF (ML\&SRF) domain, eight WPs in the Source domain, three WPs in the Damping Ring (DR) domain, two WPs in the Beam Delivery System (BDS) domain, and two WPs, in the Dump domain.  Technical preparation activities in these five accelerator domains are outlined in the following subsections. The eighteen work packages are summarized in the Appendix and detailed in the accompanying document ``Technical Preparation and Work Packages (WPs) during ILC Pre-lab''~ \cite{IDT-Tec-prep:2021}.
Estimates of the resource requirements for the technical preparation activities are presented in Subsection~\ref{sudsec:accelerator}.

Execution of Pre-lab technical preparation activities may require infrastructure that does not yet exist in every region. For instance, SRF technical preparations require facilities for cavity testing, surface treatment, conditioning of associated components, and cryomodule assembly and testing. Establishing (and funding) the necessary infrastructure is a laboratory responsibility, which may be coordinated regionally.

\subsubsubsection{Main Linacs and SRF domain}
At the heart of the ILC are the two Main Linacs (ML). They will require approximately 9,000 superconducting RF cavities assembled into  900 SRF cryomodules (CMs) and correspond to  25-30\% of the total ILC construction cost. 

SRF accelerator technology is mature. Other large-scale SRF accelerator projects exist or are being constructed, such as the European XFEL in Europe, LCLS-II in America, and SHINE in China, each on a scale of about 5-10\% of that of the ILC.
Nevertheless, SRF cost reduction R\&D continues in an international effort led by the US and Japan. Significant advances in niobium material and in cavity surface treatment have been achieved. 

Several regional hub laboratories are foreseen to be established to share the production of large numbers of ILC CMs across Europe, America, and Asia. CMs will be assembled and tested in each hub laboratory then transported to the ILC Laboratory, where final testing will be performed before installation into the ILC tunnel. This final testing will be particularly important during early production. 

Pre-lab technical preparations for the Main Linacs and Superconducting~RF domain (ML\&SRF) consist of three work packages: demonstrating SRF cavity industrial-production readiness~(WP-1), demonstrating cryomodule production readiness and global transfer while maintaining specified performance~(WP-2), and completing prototype SRF crab cavities (CCs) and completing CC cryomodule engineering design~(WP-3). 

The Science Council of Japan (SCJ) and MEXT's ILC Advisory Panel pointed out technical concerns about maintaining cavity quality during mass production and CM assembly. In response to these concerns, the Pre-lab will demonstrate industrial-production readiness for the SRF cavity and CM using cost-effective methods on a scale of roughly 1\% of the full SRF cavity production quantities.  
In WP-1, a total of 120 cavities will be produced (40 cavities per region - Europe, the Americas, Asia), and production yields will be measured in each region. 
In WP-2, six CMs (two CMs per region) will be produced and their performance tested within each region. Thus, of 120 cavities produced, 48 (40\%) will be used in the six CM assemblies. Compatibility of the CMs from different regions will be monitored.

WP-2 Pre-lab technical preparations will also demonstrate readiness for cost-effective production of other cryomodule components, such as couplers, tuners and superconducting magnets. Overall CM testing after assembling these components into the CM will be the last step in confirming CM performance as an accelerator component unit. 

America and Europe have already developed significant experience in cavity and CM production for their large SRF accelerators, including formulation of countermeasures against performance degradation after cryomodule assembly as well as during ground transport of CMs. 

As part of WP-2, resilience of CMs to intercontinental transport will be established. Preparatory work will include production of a dedicated transport cage, shock damper, and container for CM transportation. Resiliency tests will be done initially using existing European XFEL and LCLS-II cryomodules. Finally, fully tested and qualified Pre-lab CMs will be transported to Japan for repeat of QA tests and measurement at KEK, and performance before and after transport will be compared. Each CM will then return to its home region for further investigation, if necessary. 

All CMs produced by the ILC Pre-lab are expected to comply with high pressure gas safety (HPGS) regulations. Negotiation with local authorities in Japan concerning applicable HPGS regulations has already begun in preparation for the ILC Pre-lab activities.  
Upon completion of WP-2, including satisfying HPGS regulations, the CMs will be ready for industrial production during ILC construction.

SRF crab cavities (CCs) will be used in the ILC Beam Delivery System. 
CC technology has advanced considerably since the 2013 ILC TDR. High-performance crab cavities have since been developed for HL-LHC (CERN), CEBAF Upgrade (JLab), SPX (ANL), and EIC (BNL/JLab). In WP-3, the ILC Pre-lab will evaluate the now expansive array of CC technology options, select the most appropriate technology, complete CC and CC cryomodule engineering design, and produce two prototype CCs to demonstrate the technology. 
As installation of CC CMs will occur late in ILC construction, prototyping and testing of the two-cavity CC CM design will be performed early in the construction phase.

\subsubsubsection{Source domain}
Electron and positron sources produce the initial electron and positron beams. The ILC requires a high-performance polarized electron source and a high-performance positron source. The sources produce large numbers of electrons or positrons per bunch ($O(10^{10})$) with the required pattern of bunches, including high bunch repetition rate ($1.8$~MHz).
A key advantage of the ILC over circular colliders is that the beams can be polarised and that polarisation can be maintained during acceleration and collision. 
The polarised electron beam significantly improves measurement precision. Polarising the positron beam as well provides further significant improvement.

\paragraph{Electron Source} \hfill \\
Key components of the polarized electron source are the drive laser system, the high-voltage photo-gun, and the solid-state photocathodes. Since the writing of ILC Technical Design Report, the technology of all three key components has advanced. Technical preparations for the electron source~(WP-4) include work on each of these components. 

The drive laser system design and cost will be re-evaluated incorporating advances in laser technology, and a prototype system will be built to demonstrate the required beam pattern. Such a prototype demonstration was not completed before the TDR. A photo-gun with higher voltage, based on advances that have been incorporated into recent photo-guns produced for other accelerators, will be designed and prototyped. Higher voltage will relax requirements on laser pulse length and improve operating conditions for the photocathode. Strained-superlattice gallium-arsenide/gallium-arsenide-phosphide (GaAs/GaAsP) photocathodes based on recent advances will be produced to provide higher quantum efficiency and higher electron polarisation. Commercialisation of new photocathodes will be investigated. Finally, the prototype drive laser system and photo-gun will be used to test the high bunch charge, high peak current conditions from a strained superlattice GaAs/GaAsP photocathode.

\paragraph{Positron Source}
\hfill \\
Generating a polarised positron beam is more technically challenging than generating a polarised electron beam because it requires interaction of a polarised photon beam with a target and capture of the polarised positrons produced in the target. The importance of fully developing plans for the positron source was identified by both the Science Council of Japan and by the MEXT ILC Advisory Panel. A critical goal of the Pre-lab is to develop the engineering design for the polarised positron source, with credible backup plans where necessary to assure confidence.
Consequently, Pre-lab technical preparations include work on two positron source designs. The polarized positron source is based upon a polarized photon beam produced by an undulator, which is a new technique for the production of a positron beam. The backup plan is based on an electron drive beam, which is a more conventional technique for positron beam production, but which does not produce polarized positrons. (The ILC TDR defines the undulator positron source as the baseline option and the electron-driven positron source as the backup option.) 

The two positron source designs require significantly different civil engineering designs. Consequently, moving to the backup option could involve considerable cost and time if it were found to be necessary late in the project. The schedule of technical preparations has been set to enable a decision in the third year of the Pre-lab period, in order that the ILC civil engineering design can be finalised before civil construction starts. The criteria and procedure for selecting the positron source will be a responsibility of the Pre-lab. Both positron source designs will be vigorously pursued until the choice has been made. If the polarised positron source design has not adequately demonstrated technical feasibility at the time of the positron source selection, polarised positrons will be maintained as an option for a future upgrade.

The Pre-lab technical preparations for the two positron source designs are described below. Technical preparations for positron target maintenance, which are common to both designs, are also outlined.

\paragraph{\it{Undulator positron source}:}
Production of polarised positrons is based on production of a polarised photon beam by an intense electron beam passing through an undulator. The undulator is a long series of superconducting magnets, similar to undulators commonly used in light source facilities and x-ray free electron lasers (XFELs). The varying magnetic field causes electrons in the beam to radiate photons. The undulator has a helical field, unlike most light facility undulators, in order to produce circularly polarized photons. 
The length (230~m) is necessary to produce adequate intensity of polarized photons. The polarized photon beam produced by the undulator strikes a target to produce polarized positrons. The target is rotating and cooled in order to handle the heat load of the intense photon beam. The polarised positrons produced in the target must be efficiently collected by a magnetic focusing system to yield an intense polarised positron beam. Thus, key components in the polarised undulator-based positron source are the undulator, the target, and the magnetic focusing system. Three technical preparation work packages capture the necessary development by the Pre-lab for these three components.

A prototype pair of helical undulators fabricated during the TDR phase demonstrated sufficient magnetic field strength to establish the technology. Technical preparations during the Pre-lab phase (WP-5) will focus on design optimization and on simulation studies of masking to limit energy deposition in the superconducting magnets and of the impact of magnetic field errors and alignment. The target positron yield is 1.5 positrons per electron even with the 125-GeV drive beam of ILC250. (The operation of the undulator at higher energies enhances yield and facilitates operation.)

The technical challenge for the positron production target is the intense energy deposition of the incident photon beam. 
(The proposed luminosity upgrade, which would double the deposited power on the target, would increase the peak energy deposition density by 50\% and the temperature by 20\%.) 
The TDR adopted a titanium-alloy rotating target wheel with water cooling. Technical preparations (WP-6) will finalize a rotating target design based on radiative cooling, analogous to targets at some other accelerators. 
The design will be based on input from simulation and laboratory tests. 
Magnetic bearings for the rotating target wheel are part of heat load management. Technical preparations will specify requirements, explore feasibility with potential suppliers, and manufacture and test a prototype magnetic bearing. 
Finally, a full model target system will be fabricated to complement simulation studies of dynamic effects, cooling effectiveness, etc.

The magnetic focusing system, which focuses positrons produced in the target into a beam, provides optical matching. The technical challenge is achieving high positron yield. For the design in the TDR, which is based on a flux concentrator with a $3.2$~T peak field, time-dependent changes in the field will affect the positron yield. An improved design based on a pulsed solenoid is now being studied. 
Alternative designs based on a quarter wave transformer~(QWT), a plasma lens, or a new flux concentrator design will also be considered. Technical preparations of the magnetic focusing system (WP-7) will finalize the design and fabricate a prototype to be tested with the prototype target system.

\paragraph{\it{Electron-driven positron source option}:}
In the electron-driven positron source option, a driver linac produces an electron beam on a rotating target producing (unpolarized) positrons that are magnetically focused into a capture linac and then into a chicane to remove electrons, a booster linac, and into the positron damping ring. A complete technical design has been carried out; however, some further work is required to meet ILC requirements and confirm reliability. Pre-lab technical preparations will focus on three critical components: the rotating target, the magnetic focusing system, and the capture linac. 

As with the undulator positron source, the technical challenge for the production target of the electron-driven positron source is the intense energy deposition (19~kW) by the incident beam. For the electron-driven option, the target is a rotating copper disk with water channels for cooling and with a tungsten-alloy rim with which the electron beam interacts to produce positrons. Technical preparations on the target (WP-8) will include finite element method (FEM) simulation to improve heat load margin and to study target stress and fatigue, study of the lifetime of the vacuum seal necessary to maintain the required high vacuum, and target module prototyping to confirm stable operation. 

The magnetic focusing system in the electron-driven positron source option is a copper flux concentrator. The performance requirements for the ILC flux concentrator are less than those for the flux concentrator at the VEPP5 collider at BINP, Russia. Pre-lab technical preparations for the magnetic focusing system~(WP-9) will include FEM simulation of the electrical, thermal, and mechanical properties of the flux concentrator conductor and design of the power source and transmission line. System prototyping and operational tests to confirm system reliability are planned.

The capture linac design consists of an alternate periodic structure (APS) cavity in a $0.5$~T solenoid magnet. ILC performance requirements are not more stringent than those of the APS cavity in the SACLA XFEL in Japan; however, prototyping and testing should be performed to confirm stable operation and reliability. Pre-lab technical preparations for the capture linac~(WP-10) will include design and prototyping of the capture linac components, and studies of beam loading compensation and tuning method. 

\paragraph{\it{Positron target maintenance}:}
An issue common to both positron source options is radiation safety. Radiation safety both during operation and from residual activation must be considered. Radiation safety during operation can be addressed by shielding. In the case of the e-driven option, two meters of boronated concrete is expected to sufficiently limit the operating dose. The shield can be a little thinner for the undulator option due to the lower power deposition. Residual activation of the target area poses an issue for target maintenance. The highly activated target will need replacement about every two years. A remote handling system is needed for target replacement. Engineering design of the target maintenance system and fabrication of a mock-up will be performed by the Pre-lab~(WP-11).

\subsubsubsection{Damping Ring domain}
The damping rings (DRs) are circular accelerators that are placed between the electron and positron sources and the main linacs with the goal of creating high-quality electron and positron beams for the ILC. The quality of the beams is characterised by the beam \textit{emittance}. Three work packages of the ILC Pre-lab address development needed for the damping rings, focusing on system design, evaluation of collective effects, and design of injection and extraction kickers. A 2017 design change to damping ring optics in order to reduce emittance and improve luminosity at ILC250 motivates the detailed programs of the three DR work packages. The third work package also includes remaining R\&D necessary for the engineering design. The three DR work packages are outlined in the paragraphs below.

The system design of the DR beam optics will be re-optimised, following the change to the beam optics approved in 2017 to reduce horizontal emittance~(WP-12). A beam optics design that minimizes emittance tends to have smaller dynamic aperture.
Dynamic aperture of a circular accelerator is affected by the multipole errors of its magnets, especially fringe fields of the bending magnets. Thus, the design of the magnets and re-optimisation of the system accounting for those errors are necessary. Possible use of permanent magnets in the damping rings will also be investigated.

Collective effects in the damping rings with the updated DR beam optics must also be evaluated~(WP-13). Collective effects that may affect beam quality in the DRs and that need to be evaluated include impedance-driven instabilities, intra-beam scattering, space-charge effects, electron cloud effects in the positron ring, and ion effects in the electron ring. Studies based on the old TDR optics, before the reduction in horizontal emittance, found the largest sources of emittance dilution to be electron cloud instability in the positron DR and fast ion instability in the electron DR. Design, prototyping, and test of a feedback system to manage fast ion instability will also be performed, addressing a concern of MEXT’s ILC Advisory Panel.

The system design of the damping ring injection and extraction fast kickers will be updated for the EDR in order to account for updates to the dynamic aperture of the DRs~(WP-14). The injection and extraction kickers must be very fast, injecting/extracting beam bunches into/from the DRs with 6-ns spacing, and some prototyping is necessary. System design will be performed in conjunction with studies at KEK Accelerator Test Facility (ATF)\footnote{\url{https://atf.kek.jp/atfbin/view/Public/TopPageE}}, including testing long-term stability and reliability of the injection-extraction system, as called for by MEXT’s ILC Advisory Panel. 

\subsubsubsection{Beam Delivery System domain}
The ILC beam delivery system (BDS) is responsible for transporting the electron and positron beams from the ends of the main linacs, focusing them to the small sizes required to satisfy the ILC luminosity goals, causing them to collide, and finally transporting the spent beams to the main beam dumps. The ILC BDS was originally designed to cover a wide range of center-of-mass energy from 250 GeV to 1 TeV, and the TDR was written mainly for 500 GeV operation. 
Now that the ILC is to operate initially at 250 GeV, the BDS design should be re-optimized during the Pre-lab phase for 250 GeV operation, while retaining the capacity to be upgradeable to higher energies at a later time.
Two Pre-lab work packages are associated with this re-optimization. The first focuses on re-optimization of the beam optics of the final focus system, and the second focuses on re-optimization of the final doublet of superconducting magnets.
Because the design of the final focus system, particularly the final doublet, strongly impacts experiment design, BDS technical preparations will be conducted in close cooperation with the detector groups.

The final focus system (FFS) is one of the main subsystems of the BDS. Its principal purpose is to squeeze the electron and positron beams to nanometer scale at the interaction point (IP) with control of beam position to the order of a nanometer. 
The design of the ILC FFS has been validated at the ATF2 beam line at KEK, which was constructed by the international ATF collaboration for this purpose. In particular, the ability to tune the beam to achieve nanometer-scale beam size and the capability of a prototype feedback system to control the position at the IP have been demonstrated. 
During the Pre-lab phase, long-term beam stability, which was a concern in the review by the Science Council of Japan, will be tested. 
The stability test will contribute to re-optimization of the FFS design, and establish beam tuning techniques and associated hardware~(WP-15).

The superconducting final doublet magnet and cryostat packages are key components of the final focus system.  Although the technology of the final doublet was demonstrated by a series of short prototype multipole coils for the ILC TDR, superconducting magnetic coil winding technology has advanced considerably since the TDR, and new concepts for interaction region design have emerged. Consequently, the final doublet design should be re-optimized during the ILC Pre-lab phase~(WP-16). Because mechanical stability of the final doublet at the level of a few nanometers is critical to final-focus optics, vibration stability studies will also be undertaken by the Pre-lab.

\subsubsubsection{Dump domain}
Beam dumps are distributed along the ILC accelerator. Dumps are used during commissioning and tuning and continuously during regular operation. To prevent damage to the accelerator and to protect personnel, dumps receive an aborted beam in the event of an accelerator malfunction. Whereas ILC tune-up dumps are well within the performance specifications of existing accelerators, the main beam dumps for the electron and positron beams must be capable of power dissipation levels that require special consideration during design. The dump for the photon beam produced by the undulator in the polarized positron source will have high levels of localized energy deposition that also demand special consideration. Two Pre-lab work packages address the technical challenges of the main beam dumps and of the photon dump, and are described in the following two paragraphs. The technical preparations of the dumps will be led by KEK to be in accordance with local safety regulations, drawing upon experience at US and European laboratories.

The main beam dumps absorb the electron and positron beams at the end of each beam line after collision. The main beam dumps are physically large; consequently, their design impacts the ILC civil engineering design. Because the main beam dumps will experience high radiation levels that make replacement impractical, they will be designed for the full power of possible future upgrades of the ILC. For 1-TeV center-of-mass energy, the required power dissipation is 17 MW, including 20\% safety margin. The TDR design is based on the 2.2~MW water dump at SLAC. During the Pre-lab phase, a complete engineering design of a reliable, earthquake-resistant water dump system will be developed~(WP-17). Work will include studies of water flow within the dump as well as component prototyping, particularly of a robust beam window and of remote-handling facilities for replacement of the beam window. This work will address the technical concerns of the MEXT ILC Advisory Panel and Science Council of Japan reviews. 

The TDR design of the photon dump must be updated to accommodate the possible future option of ILC operation at 10-Hz collision rate for higher luminosity. The technical challenge is not the total power dissipation, which is 300 kW including 20\% margin at 10 Hz; it is the high localized energy deposition of the photon beam. Two alternative designs are under consideration, one a water-based dump, the other based on thin graphite on copper plates. Technical preparations during the Pre-lab phase will include system design and component testing of both options~(WP-18). As with the main beam dumps, the photon dump must have a reliable and safe design; however, radiation levels are significantly lower than for the main beam dumps, facilitating maintenance and replacement operations of the photon dump.

\subsubsection{Engineering design and documentation}\label{subsection:EDR}
Preparing the engineering design and documentation for the ILC accelerator is one of the principal missions of the ILC Pre-lab. Whereas the technical preparation activities described above in Subsection~\ref{subsection:WP} focus on R\&D activities that address all open technical issues or update TDR designs for significant advances in technology, engineering design and documentation activities focus on completion of a full engineering design of the ILC, including preparation of the Engineering Design Report (EDR) and all documentation necessary to initiate ILC construction. The engineering design and documentation activities will proceed in parallel with the technical preparation activities. The engineering design builds upon the TDR completed by the ILC Global Design Effort in 2013. It will incorporate the results of the technical preparation activities, as well as design changes since the TDR. 
It will also reduce uncertainty in the construction plan by scrutinizing cost and schedule risks. The engineering design and documentation activities for the ILC accelerator project will include the following items:
\begin{itemize}
\item Engineering Design Report, 
\item Engineering documentation (specifications, drawings, etc.)
\item Work Breakdown Structure~(WBS) for ILC accelerator,
\item Construction schedule,
\item Review and update of material cost estimate and human resource estimate,
\item Plans for mass production, transportation, and quality assurance, and
\item Preparation for purchase of time critical items,
\end{itemize}

Engineering design and documentation activities will be organized into work packages based on the Pre-lab WBS, which is initially being developed by the IDT and will be updated by the Pre-lab if necessary. The work will be completed as in-kind deliverables by Pre-lab member laboratories, as for the technical preparation work. A preliminary Pre-lab WBS has been used to estimate the required human resources for engineering design and documentation. This estimate is presented in Subsection~\ref{sudsec:accelerator}. No significant material resources are foreseen to be required by these activities. The next step for the IDT is to define and distribute work packages.  This step is still ongoing but less time-critical than for the technical preparation.

\subsubsection{Timeline}
The timeline for the four-year preparatory phase for two categories of activity, ``Technical preparation and production readiness'' and ``Engineering documentation'', is shown in Table~\ref{table:acc-timeline}. 
Technical preparation and production readiness activities for SRF and Positron Source are shown as examples.

\begin{longtable}{| p{.08\textwidth} | p{.41\textwidth} | p{.41\textwidth} | } 
\caption{Pre-lab timeline for technical preparation and production readiness activities, with SRF and the positron source as examples, and for engineering documentation activities.}\label{table:acc-timeline}\\
\hline 
\centering Year 
&  
\centering Technical preparation 
\newline and production readiness\newline (focusing on SRF and $e^+$ source) 
&  
\centering 
\newline Engineering documentation 
\tabularnewline %
\endfirsthead %
\caption[]{(continued)}\\
\hline 
\centering Year &\centering  Technical preparation and readiness \newline (focusing on SRF and $e^+$ source) &\centering  Final documentation 
\tabularnewline 
  \endhead
  \hline
\centering 1
&
$\bullet$ Continue cost-reduction R\&D for SRF cavities. \newline 
$\bullet$ Start pre-series production of SRF cavities in cooperation with industry. \newline
$\bullet$ Continue $e^+$ source development. \newline
& 
$\bullet$ Start review and update of TDR cost estimate by an international team.
\\ \hline
\centering 2
&
$\bullet$ Complete cost-reduction R\&D.  \newline
$\bullet$ Determine production yield. \newline 
$\bullet$ Start assembling cavities into cryomodules. \newline
$\bullet$ Review $e^+$ source designs.
& 
$\bullet$ Conduct a review on the progress for technical work and cost estimation by an internal panel. 
\\ \hline
\centering 3
&
$\bullet$ Demonstrate overseas shipment of cryomodules taking all the safety and legal aspects into account.  \newline 
$\bullet$ Select $e^+$ source design and start prototyping of critical items, e.g. $e^+$ target. \newline
& 
$\bullet$ Complete cost estimate and conduct internal and external review on the result. \newline
$\bullet$ Complete risk analysis for the technical and cost issues. \newline
$\bullet$ Complete a draft for the Engineering Design Report.  
\\ \hline
\centering 4
&
$\bullet$ Evaluate cryomodules after shipment and demonstrate the quality assurance procedure.  \newline 
$\bullet$ Establish regional organisation for the ILC component production. \newline
$\bullet$ Continue prototype work for critical components of the $e^+$ source, e.g.   $e^+$ target.
& 
$\bullet$ Complete and publish the Engineering Design Report.  \newline
$\bullet$ Start producing specification documents and drawings of large items for tendering. 
\\ \hline
\end{longtable}

\subsection{Civil construction and site-related tasks}
\subsubsection{Description of tasks and work packages}
\begin{figure}[b!]
\centering
\includegraphics[width=\textwidth]{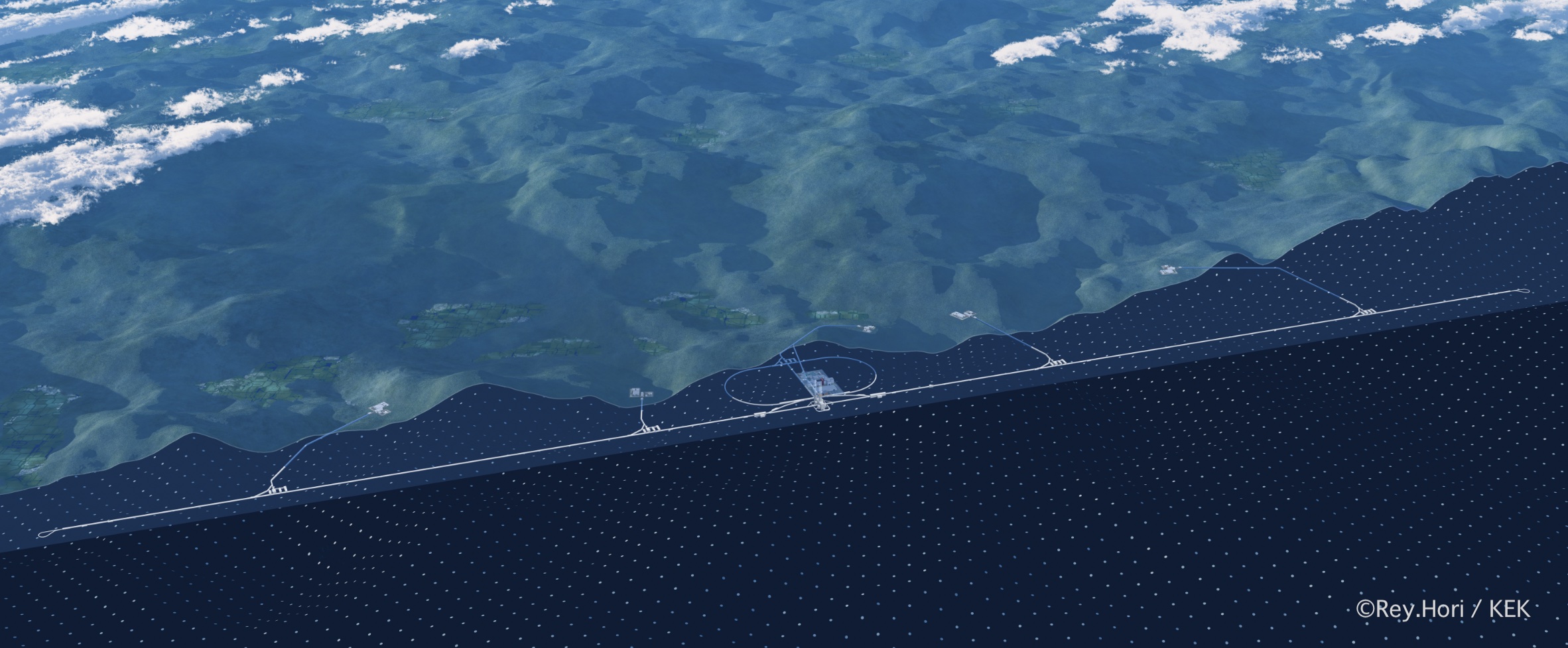}
\caption{Artist's impression of the ILC in the mountains.}\label{fig:Civil-Birdview}
\end{figure}
A candidate site in Japan was selected by an ILC community panel in Japan~\cite{Site_selection_JPN:2013} and endorsed by the Linear Collider Board~\cite{Site_selection_LCC:2013} in 2013. It is in a mountainous region \cite{Tohoku:site} (see Figure~\ref{fig:Civil-Birdview}) where granite bedrock extends  over 50 km in length, as required in the TDR.  This is more than enough for the initial 250 GeV Higgs factory which requires an accelerator length of about 20 km. Extra space will be reserved for potential future extension of the main linac for higher energy collisions. The civil engineering work will need to start at the beginning of the ILC construction phase.  Thus detailed design of civil engineering and infrastructure including the underground tunnels, underground caverns, and support facilities on the ground, must be finalized in the Pre-lab phase. 

 Large-scale underground and above-ground works for the ILC will be carried out for many years requiring national construction companies and compliance  with relevant laws and regulations in cooperation with national and local governments. Thus civil engineering works and infrastructure are considered to be the responsibility of the host country. 

\begin{figure}[t]
\centering
\includegraphics[width=\textwidth]{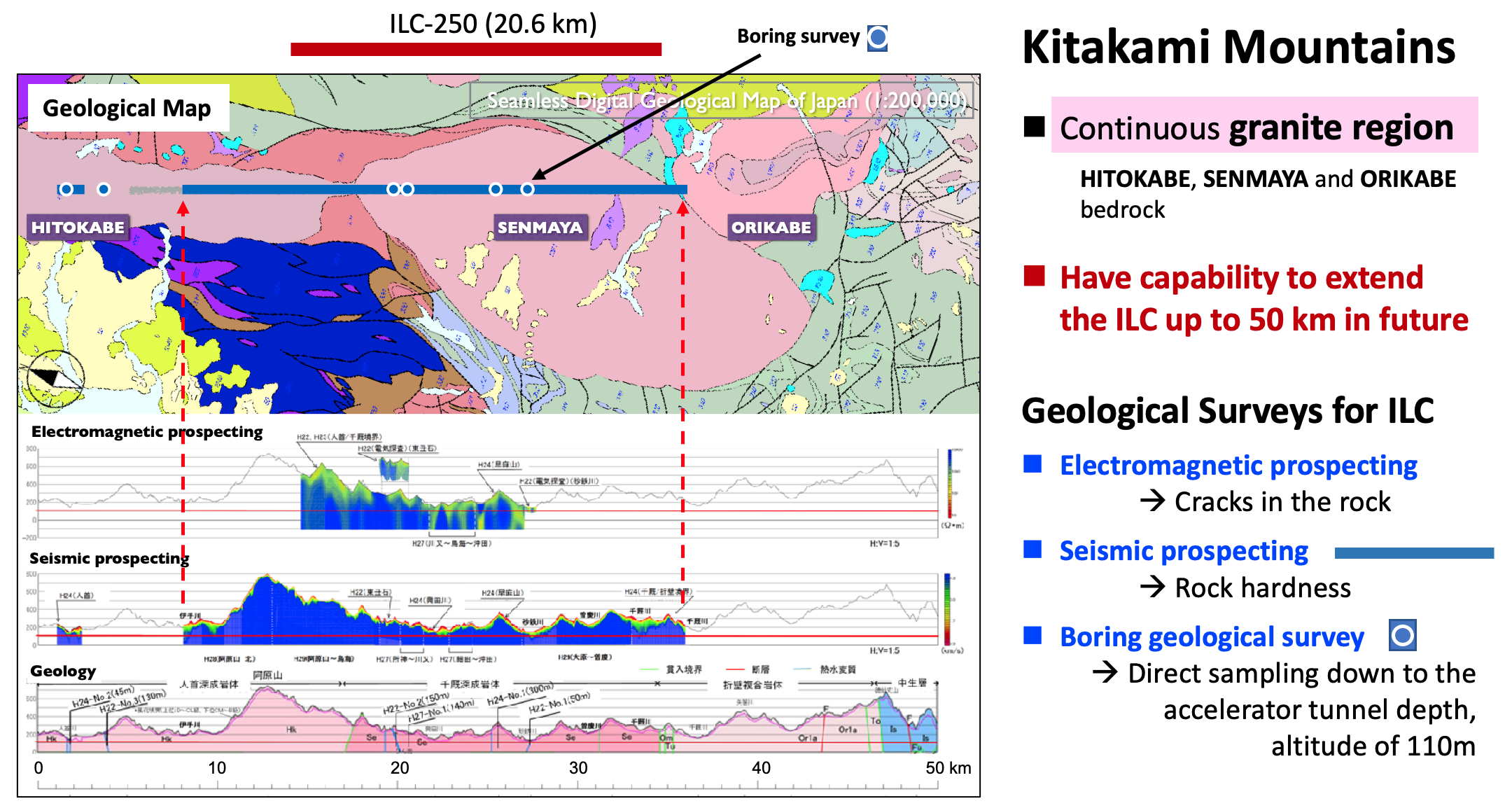}
\caption{Example of the pilot geological surveys.}\label{fig:Civil-Surveys}
\end{figure}

\paragraph{Geological surveys:}
Pilot geological surveys have been conducted at the candidate site in the Kitakami Mountains (Figure~\ref{fig:Civil-Surveys}). The straight line in the figure shows the assumed ILC route along which electromagnetic, seismic and boring survey positions that have been conducted. While the goal of these surveys was to obtain an overview of the geology of the proposed site, more detailed surveys along the accelerator route and access tunnels are required for the civil engineering work of the entire ILC system. In particular, geological issues around streams and near the surface of access tunnels require further investigation. The detector hall at the interaction point is a large underground cavern that requires an adequate structural design. Thus, it is extremely important to conduct thorough geological investigations, such as boring surveys, for the design of civil engineering works.  

\begin{figure}[tb]
\centering
\includegraphics[width=\textwidth]{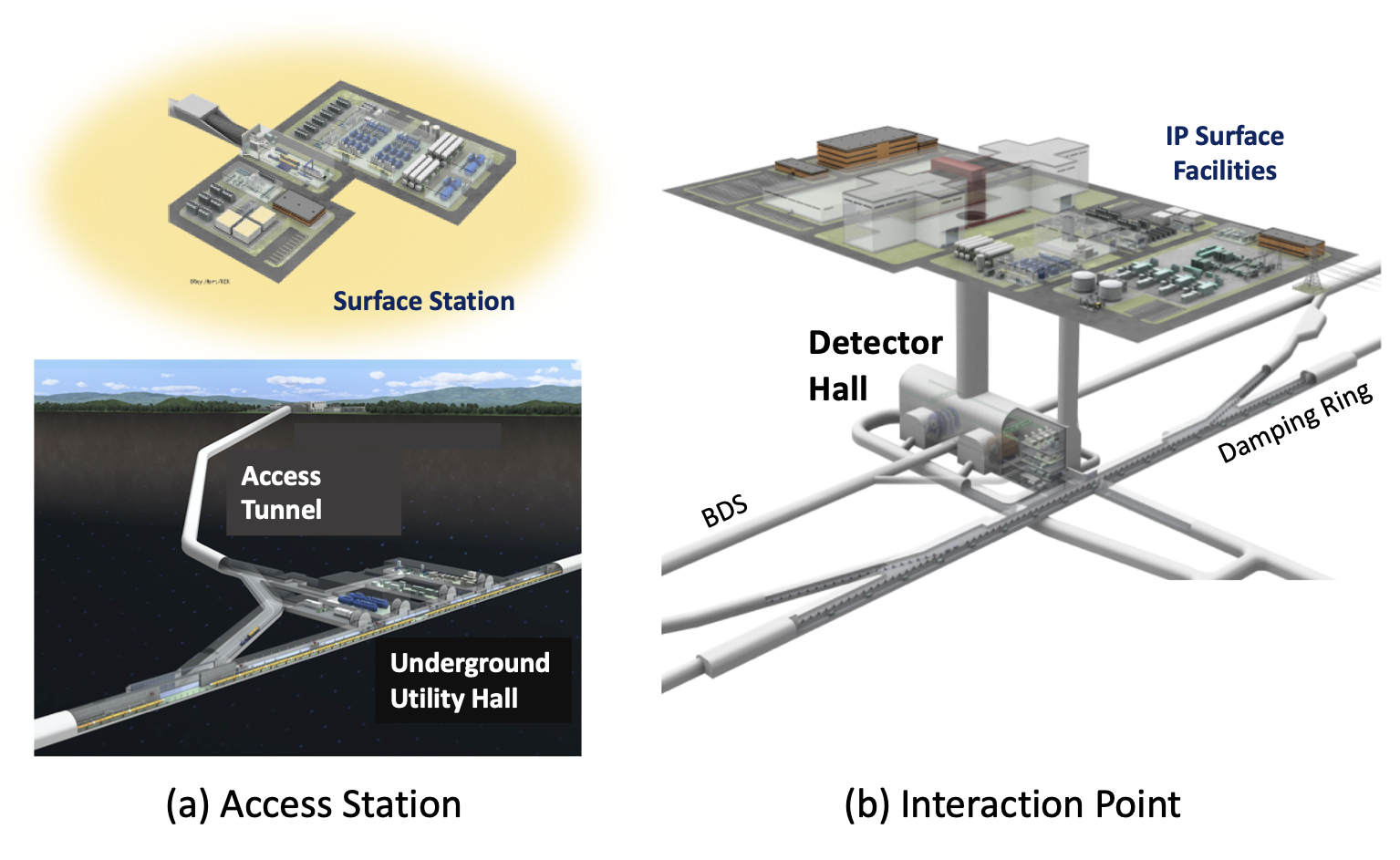}
\caption{Surface facility and linked underground structure: (a) access station, and (b) Interaction point.}\label{fig:Civil-Facility}
\end{figure}

\paragraph{Topographical surveys:}
The design of the surface facilities requires a topographical survey. The surface facilities of the ILC accelerator (see Figure~\ref{fig:Civil-Facility}) are distributed over the beam interaction point (IP) site and the five access stations - the damping ring (DR) access and four access points along the main linac. In addition, a main campus will be needed for researchers from all over the world.

The IP above-ground site has buildings for detector preparation and assembly, accelerator control, cooling water and air supply facilities that support the underground equipment around the IP, as well as the main electric power station (receiving 154 kV and distributing power to each access station). The area of IP site is  approximately 100,000 $\rm m^2$.

The access stations will be located every $\sim$5 km along the accelerator. Each will occupy an area of 19,000 $\rm m^2$ and will accommodate cryogenic systems, cooling water and air supply facilities, local power sub-stations, and a control building for personnel access.

\paragraph{Environmental assessment:}
Large scale construction will take place both underground and above-ground and environmental assessment will be an important aspect of the ILC project. The environmental assessment will be conducted in close cooperation with local authorities. A pilot survey has been conducted by the local government in the Kitakami Mountains. Future surveys will cover not only environmental but also the socio-economic impacts.  Under the Strategic Environmental Assessment, communication with local residents aim to provide a clear understanding of the assessment process. The environmental impact studies will include potential effects on air, water, handling of excavation spoil,
noise, vibration, landscape, resident comfort, and radiation. The socio-economic impacts include impacts on land use, social activities, safety, transportation, local industry and economy. Plans for temporary storage and disposal of excavation spoil are to be developed in close cooperation with local communities. Groundwater issues need to be thoroughly investigated and studied. Any potential change of groundwater level  may impact on the livelihoods of local residents.

In order to obtain external inputs on the environmental assessment, the ILC Environmental Assessment Advisory Board was established under KEK's ILC Planning Office in September 2019. The board produced a summary of discussions in December 2020 \cite{Envrionment:2020}. 

As proposed by the advisory board, staged assessment processes will be conducted during the Pre-lab phase. 

\paragraph{Safety measures:}
In designing the civil engineering and infrastructure aspects of the ILC, it is of utmost importance to establish safety measures. This is particularly important since the structure consists of long underground tunnels and underground caverns and the accelerator generates radiation. 

The main safety issues in the design of civil engineering and infrastructure are:
\begin{itemize}
\item Earthquake-proof design and fire prevention
\item Countermeasures against accidental leakage of He. (Safety design is to be adopted from large cryo-system accelerators such as XFEL and LHC.)
\item Countermeasures against power failure 
\item Design of emergency evacuation methods and escape routes
\item Management of water inflow during construction and operation
\item Storage area and disposal methods for the excavation spoil
\item Countermeasures against landslide for above-ground facilities
\item Radiation protection; design of space for radiation shields and of access control
\item Radiation confinement: controls of the air-ventilation and cooling water circulation, long-term management of activated materials especially for the beam dumps
\end{itemize}
The studies of these safety measures have been conducted starting from the early stage of the ILC project. Basic plans have been compiled in the TDR based upon the experience gained from many mountain road tunnels in Japan and large underground accelerator facilities such as the XFEL and LHC.
Robustness of safety measures will be secured by legal assessments followed by the external reviews of the final design. 

\paragraph{Detailed design of civil engineering and infrastructure:}
The basic layout of accelerator tunnels, access tunnels, utility caverns, the detector hall, and above-ground facilities   have been developed through the international and regional activities of the ILC project. The conceptual design of the underground structures was studied under international collaboration (GDE), and the design was reviewed and described in the TDR.

Civil engineering designs for underground tunnels and cavities were developed based on the standard Japanese construction methods compiled by the Ministry of Land, Infrastructure, Transport and Tourism, as well as on existing construction practices. The designs took into account information on the topography and geology of the proposed construction site at the Kitakami mountains. These basic designs are summarized in the ``Tohoku ILC Civil Engineering Plan''~\cite{Tohoku:site} along with the cost and construction schedule. The Japan Society of Civil Engineers (JSCE)\footnote{Home page in Japanese \url{http://www.rock-jsce.org/index.php?FrontPage}} assessed the technical feasibility of the designs\footnote{See (in Japanese) \url{http://www.rock-jsce.org/index.php?ILC_subcommittee_2th}} and concluded that the designs described in the "Tohoku ILC Civil Engineering Plan'' are appropriate~\cite{JSCE:report}. The JSCE also pointed out that the following three items require detailed geological investigations and design studies for the future:
\begin{itemize}
\item Parts of the tunnel that cross rivers
\item Portals of the access tunnels
\item The intersection of the top of the detector hall and the vertical shaft
\end{itemize}
The remaining tasks for the civil engineering to be carried out during the Pre-lab are to develop and document these layouts and designs for actual construction. The layout of underground facilities will be optimized by examining the land use and the environment on surface.
These processes are largely the same as the designs of other underground facilities in general such as roads, water supplies, and storage.

Work with local governments and communities to investigate the land use and the natural environment on the surface is ongoing. In the preparatory phase, choices for optimum access points on the surface suitable for construction will be narrowed, and the facility layout optimized. As the layout plan progresses, additional geological surveys and environmental assessment will be performed. The impact on construction costs and the construction period  of any large  water inflow during civil works,  will be examined and counter-measures documented. Final confirmation work will also be carried out on the transportation routes from the neighboring ports to the site, the access roads from the existing roads to the access tunnel entrances, and on the campus itself. Expert reviews will be conducted at appropriate times to complete the detailed design and the final cost evaluation.

\paragraph{Utilities:}
Utilities include the electrical power system, the cooling water system, ventilation, and air conditioning systems. The electrical and cooling water systems for the accelerator and the ventilation and air conditioning systems for the underground tunnels have been examined under the GDE, and are described in the TDR after review. Studies showed that the electrical power and water required for the operation of the ILC and the routes for receiving them can be secured. Supply routes for power and water will be optimised as the examination of the layout plan progresses.

The detailed design of a utility hall depends on the specific dimensions of the utilities, such as cooling waters, air, and power supplies, and the route for pipes and cables. The detailed design of the utilities will be carried out concurrently with the civil engineering design.

\subsubsection{Timeline}
The tasks remaining before the start of construction are as follows:
\begin{enumerate}
\item Surveys to determine detailed specifications of civil engineering facilities
\item Engineering designs
\item Preparation of material for construction contracts
\end{enumerate}
The civil engineering and infrastructure design will proceed in stages. It will start with basic designs based on the required specifications and layout plans for all underground and above-ground facilities. This will be followed by detail designs to enable construction of these facilities. 

The ILC construction period will start immediately after the preparation period (Pre-lab) is complete, and the civil engineering work will be started at the beginning of the construction period. Therefore, the detailed design required for the civil engineering work will be completed in the fourth year of the preparation period. The completion of the design will be immediately followed by preparation for construction contracts. 
Finally, detailed design should reflect the geological and topographical conditions of the site as well as the results of environmental assessments. 
Since the proposed ILC site will be located in a large area of good-quality granite, the design of the underground facilities is expected to be largely independent of the exact location of construction. However, the above-ground facilities, especially for the experimental hall and access points, should take into account the geological and topographical conditions. These specific field investigations will continue to the 3rd year of Pre-lab. 
\paragraph{Year 1:} Complete the basic plan for the layout of the accelerator and infrastructure to be documented as a basic design report. Further geological investigations and surveys will be conducted where necessary. The basic design will include safety and environmental measures. The Strategic Environmental Assessment will be initiated.
\paragraph{Years 2 and 3:} Further development of detailed design towards construction commencement. Geological and topographical surveys will be completed by the first half of the third year and the results will be incorporated into the detailed design. Progress of the Strategic Environmental Assessment will be evaluated and the environmental assessment for the Project Phase will be initiated.
\paragraph{Year 4:} Completion of the Project Phase Assessment. Complete detailed design documents. The material for construction contracts will be completed by mid-year, half a year before the end of the Pre-lab phase.%

\subsection{Preparation for physics programme}\label{section:physics}
In parallel with execution of the Pre-lab work for accelerator and civil engineering, the preparation of the physics programme will take place. An overarching goal during the Pre-lab period is to involve the worldwide particle physics community in the planning, definition and preparation of the physics programme of the ILC. In addition to the primary Higgs factory studies with collider experiments, use of single beams from injectors, stand-alone test beams or main linacs beams for more specialized experimental setups and studies will be included in the overall physics and R\&D programme.

\subsubsection{Timeline and its implementation}\label{sec:physicstimeline}
A tentative timeline has been defined starting from inviting the community to provide ideas for experiments (EoI), through evaluation and selection processes based on LoI and TP during the Pre-lab phase. The final approval of the first set of experiments at the ILC will be decided by the ILC laboratory management before entering construction. The expected process, assuming Pre-lab start up in 2022, is as follows:
\begin{itemize}
\item \textbf{2021:} The IDT calls for EoIs, to be presented in a dedicated workshop after Pre-lab start
\item \textbf{2022:} Assumed start of the Pre-lab.\newline
EoI presentations in dedicated workshop. The process of moving from EoI presentations towards LoI documents is community driven. Initial dedicated ILC R\&D funds will be needed.   
\item \textbf{2023:} LoI submissions and presentations. The ILCXAC will initiate its evaluation of the LoIs. R\&D continues. 
\item \textbf{2024:} ILCXAC recommendations of initial ILC experiments to proceed towards TPs. R\&D towards the TPs.
\item \textbf{2025:} TP submissions and presentations of these  experiments. \newline
Continuation of R\&D and recommendations by the ILCXAC based on the submitted TPs.  
\item \textbf{2026-27:} Approval of the experiments, based on the TP and ILCXAC recommendations, by a committee set up by the ILC Laboratory. Recommendations to proceed towards Technical Design (TDR) Reports. Funding requests for construction are being prepared and submitted according to the relevant procedures for the participating institutes. 
\item \textbf{2027:} The ILC laboratory allows construction to start and construction funding spending for experiments or experimental subsystems based on TDRs approvals. 
\end{itemize}

The steps above will be based on common guidelines for experimental schedules, cost-books and estimates, resource estimates, common funds concepts, central laboratory versus experimental responsibilities, and selection criteria and procedures. 

Beyond the initial experiments at ILC, additional or future experimental proposals will follow a similar path and evaluation procedures.

\subsubsection{Coordinated activities}
Beyond implementation of the steps in the timeline above, several centrally coordinated processes and activities are needed in order to
prepare the physics programme. These are either common challenges for all the experimental groupings during the Pre-lab phase, or concern the interfaces between the physics programme and the accelerator, civil engineering and laboratory infrastructure. Typical examples of the former are detector R\&D, organising access to test beam and irradiation facilities worldwide, and planning and development of software and computing infrastructure. Examples of the latter are definition and optimisation of the interfaces between the  detectors and the accelerator, laboratory space and logistics planning, definition of power and cooling needs, safety rules and regulations, etc. The tasks above fall under the responsibility of Associate Director for Research (see Subsection~\ref{subsection-research-dir}).

Other more general activities related to preparation of the physics programme are workshops, theory programmes and planning of facilities and services for the user community. Dissemination and outreach will be an important activity where communicating the physics programme will play a crucial role.
\section{Reference cost and required human resources}\label{section:cost}
\subsection{Accelerator}
\label{sudsec:accelerator}
Material cost\footnote{Material cost includes also cost for fabrication, transportation, and contracted engineering work.} and personnel requirements for in-kind deliverables in the technical and engineering work have been estimated based on the WBS tables under development by the IDT. %

Resource requirement estimates for the deliverables of the eighteen technical preparation work packages in the six accelerator domains described in Subsection~\ref{subsection:WP} are summarized in Table~\ref{table:acc-resources}: a total of 58 MILCUs~\footnote{ILCU: The ILC currency unit, where one ILCU is equal to one 2012 USD.} for material costs and 364 full-time-equivalent years (FTE-yr) for personnel. Additional resources will be required to develop and operate infrastructure needed for the technical preparations, such as for SRF development and for the operations of the Accelerator Test Facility (ATF). Required resources depend on the region (laboratories) where the work will be executed and are therefore not included in the estimation here. 

In order to be ready for the start of ILC construction, engineering design documentation, such as specifications and drawings, must be prepared. Documentation needs to cover all ILC accelerator areas, including Accelerator/Engineering Design and Integration (ADI/EDI), electron and positron Sources, Damping Ring (DR), beam transfer system from the DR to the Main Linacs (RTML), Main Linac (ML), and Beam Delivery System (BDS).
Table~\ref{table:acc-human} summarises the human resource requirements for the completion of the engineering design documentation and construction preparation work: a total of 250 FTE-yr, where the work will be shared among the participating laboratories, as described in Subsection~\ref{subsection:EDR}, and coordinated by the Central Technical Office.  

\begin{table}[tb]
\centering
\caption{List of estimated material costs and human resource requirements for deliverables of the technical preparation activities, where ILCU is defined in the text. (Resources for the infrastructure needed for deliverables are not included.)}\label{table:acc-resources}
\vspace{2mm}
\begin{tabular}{|l|r|r|}
\hline
Domains   & \multicolumn{1}{|c|}{Material cost}      & \multicolumn{1}{|c|}{Human resources}   \\
      &    \multicolumn{1}{|c|}{[MILCU]}   & \multicolumn{1}{|c|}{[FTE-yr]}          \\
\hline \hline
Main Linacs (ML) and SRF &  41.25 & 285 \\
\hline
Electron Source  & 2.60 & 6\\ 	
\hline	
Positron Source  & 5.85  & 15 \\	
\hline
Damping Ring (DR) 	 & 2.50 & 30\\
\hline
Beam Delivery System 	 & 2.20 & 16 \\
\hline
Dump    & 3.20 & 12 \\
\hline
\hline
Total      & 57.60 & 364
 \\ \hline
\end{tabular}
\end{table}
\begin{table}[b!]
\centering
\caption{Estimated human resource requirements for engineering design and documentation.}\label{table:acc-human}
\vspace{2mm}
\begin{tabular}{|l|r|}
\hline
Item   &  \multicolumn{1}{|c|}{Human resources }   \\
      &    \multicolumn{1}{|c|}{[FTE-yr]}   \\
\hline \hline
Accelerator/Engineering design and integration & 75 \\
\hline 
Sources  & 35\\ 	
\hline
Damping Ring (DR)	 & 30\\
\hline
Beam transfer system from DR to ML
& 25\\
\hline 
Main Linacs (ML)  & 60 \\
\hline
Beam Delivery System 	 & 25 \\
\hline \hline

Total      & 250
 \\ \hline
\end{tabular}
\end{table}
In summary, the material cost for Pre-lab accelerator work is estimated to be 58 MILCUs and human resource requirements are estimated to be $\sim 620$ FTE-yr. As already indicated, resources for infrastructure to execute WP's are not included. The indicated numbers should be considered as reference values. Final numbers will depend on the deliverables and specific requirements of the laboratories that assume responsibility for them, and will be re-evaluated later to account for actual laboratory contributions. Detailed breakdowns, needed for the laboratories interested in taking responsibilities to plan their budgets, will become available during the forthcoming discussion of the start-up process of the Pre-lab.

Most ILC accelerator components will be provided through in-kind contributions. It is expected~\cite{KEK:action_plan} that roughly 260 people will be working on accelerator-related activities worldwide during the first year of ILC construction. Although much experience exists in the construction of large-scale accelerators, it is important to ensure that experts continue to be trained. On-the-job training of personnel through execution of the technical preparation WPs and the development of the associated infrastructure during the Pre-lab phase will enable an effective transition to the construction phase. The envisioned number of about 250 accelerator-related personnel engaged in the fourth year of the Pre-lab phase in the laboratories (a part of the above estimate) will be essential for this smooth transition.  

\subsection{Civil engineering and site-related activities}
Civil engineering work must comply with the local regulations and constraints. Therefore, the host country will be responsible for civil engineering and other construction site-related activities of the Pre-lab. It is estimated \cite{Tohoku:site} that approximately 22 MILCU will be required for site-related studies such as geological survey, land survey for surface facilities, environmental assessment, etc. It is estimated that 43 MILCU will be required for the detailed design and documentation of civil engineering and infrastructure facilities in order to start ILC construction immediately after completion of the ILC Pre-lab. Most of this work will be outsourced to the private sector, including design, survey, and construction companies. It is anticipated that approximately 70 FTE-yr would be needed to manage these various tasks. Table~\ref{table:site-resources} summarizes the estimated resource requirements.
\begin{table}[h]
\centering
\caption{Estimated civil engineering cost and human resources requirement.}\label{table:site-resources}
\vspace{2mm}
\begin{tabular}{|l|c|c|}
\hline
Item   & Cost      & Human resources   \\
       & [MILCU]   & [FTE-yr]          \\
\hline 

Site surveys      & 22    & \multirow{2}{*}{70}\\
Detailed designs  & 43    & \\
\hline 
\end{tabular}
\end{table}
\subsection{Central Bureau}
The Central Bureau will be established in Japan, where most of its staff will be located. Table~\ref{tab:central-bureau-human} summarises the current estimate for the required human resources of the Central Bureau\footnote{Based on  an estimate made by KEK.}, totalling 30 FTE per year. In addition, an annual operation budget of 820 kUSD is estimated in order to cover office rent, travel costs for the Central Bureau and committee members, conference and workshop contributions, outreach and public relations, external consultants, and other miscellaneous items. In current thinking, the cost of the administrative staff and operation would be covered by the host country, Japan. The remaining human resources would be provided through cash contributions or by in-kind temporary relocation to Japan of people with the necessary skills. A cash common fund will be needed to pay the salaries of the director and associate directors in order to avoid appearance of ``conflict of interest''.  
\begin{table}[bth]
    \centering
    \caption{Pre-lab Central Bureau human resource requirement}    \begin{tabular}{|l|c|}
    \hline
        Item & FTE/Year \\ 
        \hline
        Directorate Office & 12 \\
        \hline 
        \hspace{2mm} Director and associate directors & 4 \\
        \hspace{2mm} Secretarial support, legal service, communication, safety & 8
        \\        \hline
        Administration Office & 9 \\
        \hline 
        \hspace{2mm} Head & 1 \\
        \hspace{2mm} International Relation, Finance \& Procurement, &8 \\ 
        \hspace{6mm}Human Resources \& Travel, Local IT service & 
        \\        \hline
        Central Technical Office & 9 \\
        \hline 
        \hspace{2mm} Project management and technical coordination & 5 \\
        \hspace{2mm} Coordination for the common physics and detector needs & 2 \\ 
        \hspace{2mm} IT service for Engineering Data Management System & 2
        \\        \hline
        \hline
        Total & 30 \\
        \hline
        
    \end{tabular}

    \label{tab:central-bureau-human}
\end{table}

\newpage

\vfill 
\newpage
\appendix
\section{Appendix}
This Appendix provides additional detail concerning each of the technical preparation work packages outlined in Subsection~\ref{subsection:WP}.
Further detail is provided in the document ``Technical Preparation and Work Packages (WPs) during ILC Pre-lab''~ \cite{IDT-Tec-prep:2021}.

\subsection{Main Linacs and SRF domain}
\subsubsection{Work Package 1 (SRF cavity industrial-production readiness)}
WP-1 has the critical and central aim to prepare for and demonstrate SRF cavity industrial production readiness. The plan is based on global fabrication of 120 cavities (40 from each of the three regions - Europe, the Americas and Asia). From 120 produced, 48 cavities will be used in WP-2 to produce six CMs, two from each region. The remaining cavities will provide sufficient statistics to test the requirement that $\ge$ 90\% pass the RF performance criteria, which are $\rm 35~MV/m$ at $Q\ge 0.8\times 10^{10}$ and $\rm 31.5~(\pm 20\%)~MV/m$ at $Q\ge 10^{10}$. 
Cavity production includes the cavities, helium tank, magnetic shield, surface treatment, satisfaction of high-pressure-gas safety (HPGS) regulations\footnote{Negotiation of HPGS regulations with local authorities in Japan has already begun in preparation for ILC Pre-lab SRF activities for WP-1 and WP-2.}, and second and subsequent vertical tests if required.

Cost-effective cavity production including niobium material and surface treatment recipes/methods is needed.
Prior to production of cavities, detailed specifications will be established to ensure compatibility across production centres.
Production processes including surface treatment will be standardized for yield evaluation. 

 Each region will require significant infrastructure  and technology development for cavity production, such as electron beam welding machines, vertical cryostats, surface treatment facilities, vacuum furnaces for heat treatment, and pre-tuning machines. This infrastructure is expected to be provided by laboratories in each participating region. The cost of the required regional infrastructure and its development is a regional responsibility.

\subsubsection{Work Package 2 (Cryomodule assembly and transfer)}
The cryomodule (CM) assembly and transfer work package will demonstrate CM and CM component production readiness. It consists of: production of CM components including vacuum vessel; production, assembly, and test of complete CMs in each region; and validation of intercontinental overseas shipment. 

The major CM components to be produced are couplers, tuners, and superconducting combined function magnets.
These components must be capable of reliable long-term operation, and capable of future cavity performance upgrades.
The superconducting magnets must withstand SRF dark current irradiation and heating arising from high-gradient SRF linac operation. 
Irradiation and heating effects will be mitigated by an absorber that minimizes heating of the coil and/or use of a superconductor with higher critical temperatures, such as Nb$_3$Sn/MgB$_2$.

Two complete cryomodules will be produced in each of the three regions, Asia, the Americas, and Europe. 
The CMs produced will be Type B, which has a superconducting magnet at its center.
The first quality assurance (QA) tests of the six assembled Pre-lab CMs will be carried out in each region before international transport. The Pre-lab CMs must also satisfy the local Japanese high-pressure-gas safety regulations that are negotiated for the ILC.

Although previous projects, including the European XFEL and LCLS-II, confirmed that  cavity performance remained acceptable after CM ground transportation, an ILC-type cryomodule has never been shipped by sea. This WP will realize the first overseas shipment to demonstrate overall SRF technology readiness for global ILC production. Safe transportation will require development of a dedicated cage, shock damper, and container. Transportation tests will be performed in two stages. In the first stage, fully constructed CMs from LCLS-II or the European XFEL that are not fully suitable for use in a linac will be used to gather important information about mechanical stress during transportation. In the second stage, prototype ILC CMs will be shipped to Japan after QA testing.  
One CM each from the Americas and Europe will be transported to Japan by overseas shipping. QA tests will be performed again at KEK after shipping, in order to re-measure performance and verify that quality requirements are still satisfied. Following testing, each CM will return to its home region for further investigation, if necessary. 

Each region will require infrastructure and utilities for cavity and CM testing. As for WP-1, the cost of required regional infrastructure, including its development, is a regional responsibility.

\subsubsection{Work Package 3 (SRF crab cavities for BDS)}
SRF crab cavities (CCs) near the ILC interaction point (IP) are crucial to achieving the highest possible luminosity.
The CCs need to be installed in a constrained space in the Beam Delivery System near the IP. As presented in the ILC TDR in 2013, the baseline CC technology choice proposed was a 9-cell, 3.9 GHz elliptical design, incorporating a lateral deformation in order to enable appropriate separation of the operating dipole $\pi$-mode frequency orientations. 
Since 2013, there have been extensive CC developments undertaken for other accelerators, utilizing a variety of alternative technology solutions that can provide compact integration into constrained accelerator environments, whilst retaining strong HOM damping and kick voltage performance.

The proposed scope of the CC system development programme for the ILC Pre-lab phase is to complete an assessment of candidate technologies, selecting a demonstrated CC technology based on prototype CC development and performance evaluation of two of the most optimum technology options. The final down-selected technology will be used as the basis for developing a complete engineering design of a two-cavity prototype cryomodule which meets all ILC IP implementation constraints.

It is anticipated that the collaborating groups across Europe and America will start their respective CC technology studies with electromagnetic design ahead of the Pre-lab phase. During Pre-lab Year-1, an evaluation of the potential CC design options, including mode couplers and tuners, will be performed with the aim of down-selecting the two most optimum integrated designs. These two designs will then be taken forward to prototyping and validation in Year-2. 
The final technology down-selection process will be made based on high-power evaluation of the prototypes. 
Preliminary design of an integrated two-CC cryomodule will be completed in Year-3.
In addition in Year-3, the two prototype CCs will be configured in a vertical cryostat in order to provide a provisional assessment of the timing and phase synchronization performance of the pair at expected ILC gradients.
The final engineering design of the CC CM, including its interface to the ILC beam line is anticipated to be established by the end of the Pre-lab phase in Year-4.
A prototype full two-CC CM of the final design will be manufactured and tested in a few years period at the beginning of the ILC construction phase, in plenty of time for final CC CM production and installation. 
\subsection{Source domain}
\subsubsection{Work Package 4 (Electron source)}
The baseline design of the polarized electron source, includes the drive laser, a 200 kV DC high voltage photo-gun, GaAs/GaAsP photocathodes which provide polarization $>$85\%, and the design requirements of the electron injector. While there are no foreseeable “show-stoppers” leading to the construction of the ILC polarized electron source, there remain unfinished critical technical tasks from the GDE period which include completing a prototype drive laser, and then using it to test the high bunch charge, high peak current conditions from a strained superlattice GaAs/GaAsP photocathode from the high voltage gun.  Additionally, since the GDE there have been meaningful technological improvements in lasers, high voltage guns and photocathodes which should be incorporated to the baseline design and incorporated, as opportunities for reliability, performance or cost improvement. 
\subsubsection{Work Package 5 (Undulator)}
In the undulator-based positron source, polarized electrons are produced by circularly polarized photons incident on a target. The polarized photons are produced by a high-energy electron beam traversing a series of superconducting helical undulators 231 m in length, optimized for 125 GeV drive beam energy when ILC is operating at 250 GeV center-of-mass energy. Each undulator has a field length of 1.75 m. Two undulators are mounted in each cryostat, which operate at 4.2K. 
A pair of prototype undulators exhibited sufficient magnetic field strength during the GDE phase. The principal design considerations to be addressed by WP-5 during the Pre-lab phase are alignment, masking, and 125 GeV operation. Experience with alignment, both with and without beam, of long undulators operating at other facilities, for instance XFEL, will be incorporated into undulator design studies planned for the Pre-lab phase. The impact of misalignment and of undulator field errors will be studied by detailed simulation. Masking will protect the undulators from excessive heating by photons radiated by the electron drive beam. Masking requirements and design will also be studied via detailed simulation. Finally, further optimization of undulator parameters to enhance the photon beam for the initial drive beam energy of 125 GeV will be performed via simulation and engineering studies.

\subsubsection{Work Package 6 (Rotating target for undulator scheme)}
The TDR adopted a target made of a titanium alloy (Ti-6Al-4V) of 14 mm thick (0.4 radiation length). The target is mounted at the rim of a wheel with a diameter of 1 m and rotating at 2,000 rpm (100 m/s at the rim). An ILC prototype target wheel of an early model has been constructed and commissioned. This wheel is placed in a vacuum of ~$\sim10^{-6}$ Pa. In the current ILC250 design, the target thickness is reduced to 7 mm without any yield loss. The heat deposited by the beam is approximately 2 kW. 
The main problem encountered in previous studies was cooling. The TDR adopted a water-cooling system, with magnetic fluid as the vacuum seal; however, the R\&D on this system was discontinued because of vacuum leakage through the seal. Since then, a target with the radiation cooling mechanism has been investigated. To date, principal engineering studies have been conducted, but detailed engineering and manufacturing studies have not yet been performed. 

WP-6 is focused on the target model, from design finalization to fabrication of a full model.
Radiation cooling is a promising new concept for the ILC positron target. There are already several prototype examples in former experiments, where radiation cooling has been used. For instance the graphite target at CNGS (CERN), immersed in stationary He gas, was cooled mainly by radiation complemented by natural convection. As well, experiments at FRIB-US, J-PARC, PSI and RAL-UK have studied or used radiation cooled targets.

\subsubsection{Work Package 7 (Magnetic focusing for undulator scheme)}
The magnetic focusing system, which focuses positrons produced in the target into a beam, provides optical matching. The technical challenge is achieving high positron yield. In the TDR, a flux concentrator with a 3.2 T peak field as the optical matching device (OMD) was adopted. It was expected to have a field flat-top of approximately 1 ms; however, it was subsequently found that time variation of the field is inevitable for such a long pulse due to the skin depth effect. 
An improved design based on a pulsed solenoid is now being studied. 
Alternative designs based on a quarter wave transformer~(QWT), a plasma lens, or a new flux concentrator design will also be considered. Technical preparations of the magnetic focusing system (WP-7) will finalize the design and fabricate a prototype to be tested with the prototype target system.
\subsubsection{Work Package 8 (Rotating target for e-driven scheme)}
In the e-driven scheme, the positron production target material is a tungsten-rhenium (W-Re) alloy. A W-Re rim with a diameter of 0.5 m and a thickness of 16 mm is rotated in vacuum with a tangential speed of 5 m/s (225 rpm). The W-Re rim is attached to a copper disk fixed to a rotating shaft with water channels for cooling. The shaft is supported by a couple of mechanical bearings, and the vacuum is sealed by ferrofluid. The ferrofluid seal is an organic solvent with fine iron powder that fills the gap between the rotating shaft and unit body to create the seal; it is held in place by a permanent magnet. The motor, bearing, and rotatory joint for the water inlet are exposed to air.

WP-8 aims for more accurate calculation of target stress and fatigue to improve the target design. The target will be highly activated by the beam. Therefore, confirmation of reliable long-term operation with a prototype is an essential part of the WP.

\subsubsection{Work Package 9 (Magnetic focusing for e-driven scheme)}
A two-conductor flux concentrator made of copper is considered for magnetic focusing in the e-driven scheme. The primary conductor is a spiral coil; it generates a B field along the axis. The other component is the secondary conductor. The eddy current in the secondary conductor, which is induced by the primary B field, flows and generates a B field in the conical space. The target is placed outside of the smallest aperture where the B field is strongest. A 5 T field is induced along the axis. The diameter of the beam hole is 16 mm. 

WP-9 aims to complete the engineering design of the flux concentrator system, including the thermal and electrical design of the conductor, the transmission line design, and the power source design. System prototyping is required to confirm reliable operation. 

\subsubsection{Work Package 10 (Capture cavity for e-driven scheme)}
The capture linac for the e-driven positron source scheme consists of an L-band alternate periodic structure (APS) cavity surrounded by 0.5 T solenoid magnets. The foremost reason for its structural form is its wide aperture, which affords better RF stability than the $\pi$-mode standing wave cavity. 
For a klystron, it requires a 50 MW power supply with a 2 $\mu$s pulse width. Although there is no commercially available klystron that satisfies these requirements, an S-band klystron that has better performance exists.

WP-10 consists of RF design and prototyping of the APS cavity for the capture linac, prototyping of the cavity power unit, prototyping of the solenoid, and test operation of the cavity and power source. Because the operational mode of the capture linac is unique (because it uses the deceleration capture method), beam loading compensation and tuning methods must be studied. The power unit prototype requires fabrication of an L-band klystron by scaling the existing S-band klystron design. The operational test of a capture linac system of prototype components will confirm required high system reliability.

\subsubsection{Work Package 11 (Target replacement)}
Radiation from the positron source target is confined by a 2~m thick boronized concrete shield. After 100 hours of cooling, a 10 Sv/h dose is still expected on the target surface. The target must be replaced every two years due to radiation damage. During replacement, radiation exposure to workers must be well controlled. The target module consists  of the target, flux concentrator, first acceleration cavity, etc. Many of the joint connections for RF, electric power, water, control, etc. are assembled on the front panel of the module, and these joints should be safely disconnected. WP-11 aims to complete the technical design of the target replacement system, to fabricate a mock-up, and to develop a fail-safe system.
\subsection{Damping Ring domain}
\subsubsection{Work Package 12 (System design)
}
The dynamic aperture of the ILC damping rings was evaluated by assuming hard-edge ideal magnets; however, the dynamic aperture of a circular accelerator is affected by the multipole errors of its magnets, especially the fringe fields of the bending magnets. Therefore, WP-12 includes DR magnet design and re-evaluation of DR beam optics considering the multipole errors of the actually designed DR magnets. After evaluation of the dynamic aperture with the current beam optics, the DR lattice will be further optimized to improve the horizontal emittance while maintaining the dynamic aperture tolerance.

In addition, WP-12 includes investigating the feasibility for introducing permanent magnets (PM) in the arc sections of the DRs. One of the major advantages of PMs is reduced power consumption and operating costs relative to electromagnets. Other benefits include reduced infrastructure (no large power supplies or water pipes), and lower vibrations (no flowing water). Disadvantages are that PMs are fixed-field, sensitive to small changes in temperature, and susceptible to radiation damage. It is necessary to investigate PM magnetic field uniformity, stability, and radiation damage by prototyping several field-adjustable PMs during the ILC Pre-lab period. The decision whether to use PMs will be made during the Pre-lab period, taking into account a wide range of factors, including not only results of PM prototyping, but also experience during Pre-Lab period with PMs used in 4th-generation light sources.

\subsubsection{Work Package 13 (Collective effect)}
WP-13 covers simulations of collective effects in the damping rings following the change made to damping ring optics in 2017 for higher luminosity. Before this change, the largest sources of emittance dilution were identified to be electron cloud (EC) instabilities in the positron DR, and fast ion instabilities (FII) in the electron DR. The studies will focus on these two effects, but also include the effect of ion-trapping instabilities. 

WP-13 will in addition perform system design, including beam tests, for a high-resolution fast feedback system to control fast ion instabilities. This work will be based on the experience and upgrades of SuperKEKB, which has a circumference close to that of the ILC DRs and a feedback system similar to ILC250. If simulations indicate that additional experimental studies in FII suppression  is needed under conditions different from those at SuperKEKB, beam tests should be performed at other accelerators. 

\subsubsection{Work Package 14 (Injection/extraction)
}
Considering the current dynamic aperture of the present design of the ILC DR, the electrode gap of the stripline kicker must be increased. It is also necessary to make minor modifications to the optics in the straight section of the DR as well as in the injection and extraction lines. A long-term stability test of the fast kicker system will be performed at the ATF. The kicker pulser used for the long-term test will be the drift step recovery diode (DSRD) pulser in ATF. The power that can be supplied by the DSRD pulser is however limited and there is no margin when applying it to the ILC. We will therefore in parallel develop a power source that is capable of realizing higher voltage.  Furthermore, because the injection system for the electron-driven position source is different from other ILC injection and extraction kickers, a specific injection kicker for this purpose need to be developed.
\subsection{Beam Delivery System domain}
\subsubsection{Work Package 15 (Final focus)}
The beam size at the ATF2 focal point is designed to be 37 nm, which is technically equivalent to a 7.7 nm beam size for ILC250. A vertical electron beam size of 41 nm, which essentially satisfies the ATF2 design goal, has been obtained at ATF2, with a bunch population of approximately 10\% of the nominal value of $10^{10}$ electrons and with a reduced aberration optics. Recent studies indicate that the vertical beam size growth with the beam intensity due to wakeﬁeld effects. Furthermore, there are also technical concerns about the technology of the control and feedback systems and the long-term stability of the beam focus and position in the ATF2 beam experiment. To address these concerns further ATF studies are foreseen during the Pre-lab period. In parallel, the ILC final focus system (FFS) design will be assessed from the point of view of beam dynamics, choice of technology and hardware, and long-term stability operation issues. To implement an experimental program based on the already unique results achieved by the ATF/ATF2 collaboration, an ATF3 collaboration is underway with the existing ATF2 partners plus new possible members worldwide. The ATF3 results are expected to provide important information necessary for the system design of the ILC final focus beam line. Through these studies we will optimize the FFS design for the initial  ILC energy of 250 GeV, also taking into account compatibility with future energy upgrades.

\subsubsection{Work Package 16 (Final doublet)}
The final doublet (FD) produces the small beam size at the collision point and consists of two superconducting quadrupole magnets (QD0 and QF1). Superconducting coil winding technology has advanced greatly since the TDR was finalized, and later projects have proposed and/or implemented new interaction region (IR) design options. WP-16 includes re-optimization of the FD design to incorporate advances in technology and beam optics design. 

As input to the FD design, WP-16 will also study the stability of the QD0 magnet against vibration induced by its cryogenic system. ILC luminosity will be very sensitive to the stability of the QD0 magnet. Vertical vibration of QD0 must not exceed around 50 nm in order to stay within the capture range of the intra-train collision feedback system. This requirement is beyond the experience from existing accelerators. For this reason, superfluid helium cooling for QD0 was chosen in order to minimize vibration of QD0. QD0 vibration arising from the cooling system will be evaluated during the Pre-lab period.

Because the FD design is strongly connected to the detector designs, the technical preparation will be done in close cooperation with the detector groups.
\subsection{Dump domain}
\subsubsection{Work Package 17 (Main dump)}
This WP covers further design of the main beam dump, demonstration of the stability of its window and the handling procedures for the window.

The design work will be carried out in collaboration with experts from the field of high-power targets and dumps worldwide. CERN operates beam dumps for large accelerators and high-power beam dumps, and SLAC and JLAB have experience with water-circulated beam dumps. KEK will lead the system design of the beam dump facilities, ensuring environmental and radiation safety in collaboration with governmental bodies, industry, and the scientific community. The engineering design of the vortex flow system in the water dump vessel and the overall water circulation system will be done following the experiences at SLAC and JLAB. The stability of the window will be confirmed from the perspective of radiation damage and mechanical robustness. The Ti alloy, Ti-6Al-4V, was selected as a window material following the experience globally with high-power targets and dumps, mostly for proton beams. Further collaborative studies towards increasing the robustness will continue. The mechanical robustness of the window will be confirmed through sealing prototypes and demonstration of remote exchange for maintenance work in a high radiation environment. A scheme for monitoring the integrity of the window will also be studied. The design for safety, that is, earthquake protection, containment of activated water, including countermeasures for failures, is a major engineering issue to be addressed. A maintenance plan will be presented with a concrete design of the handling equipment for the dump system. These studies will be conducted in collaboration with industries.

\subsubsection{Work Package 18 (Photon dump)}
The photon dump for the undulator positron source will absorb an average power of 120 kW for the 250-GeV  high-luminosity case. For the possible future option of 10-Hz collisions, a 300 kW compatible design is needed, including a 20\% safety margin. The dump design needs to be changed from TDR, where a water dump similar to the main beam dump was assumed. 

Two designs are currently proposed. One is a water-based dump and the other is a graphite-based dump, both of which will be installed 2 km downstream of the positron target to reduce the photon load density. The 2 km photon transport line passes next to the BDS and shares the BDS tunnel. The dump can be installed with appropriate shielding in the space created at the junction of the RTL (Ring to Main linac) and the BDS beam lines. These designs are based on heat and radiation damage analyses, and need to move forward by incorporating technical challenges, especially the power absorption structures and the maintenance of activated equipment.

These design studies will be carried out in collaboration with experts from the field of high-power beam targets and beam dumps throughout the world, as well as those with experience in high-power photon absorbers for XFELs and fourth-generation light sources. Prototyping of the key structures is foreseen as part of the WP.

\end{document}